\documentclass[twocolumn,a4paper,10pt]{extarticle}
\usepackage{subfig}
\usepackage{mathptmx}

\usepackage[compact]{titlesec}

\titleformat{\section}[block]
{\fontsize{14}{15}\bfseries}
{\thesection}
{1em}
{}
\titleformat{\subsection}[block]
{\fontsize{11}{15}\bfseries}
{\thesubsection}
{1em}
{}

\usepackage[margin=0.78in]{geometry}

\usepackage{amsmath}
\usepackage{newfloat}
\usepackage{listings}
\usepackage{balance}

\usepackage[final]{microtype}
\usepackage[sort,nocompress]{cite}

\usepackage{graphicx}

\usepackage{algorithmic}

\usepackage{booktabs,tabularx}

\usepackage[hyphens]{url}
\usepackage{hyperref}

\usepackage{mathtools}
\usepackage{centernot}

\def\name{GhostMinion}

\usepackage{cleveref}
\usepackage{framed} 
\usepackage{amssymb,amsmath,mathtools, amsthm}
\usepackage[framemethod=tikz]{mdframed}

\lstdefinelanguage{JavaScript}{
  keywords={typeof, new, true, false, catch, function, return, null, catch, switch, var, if, in, while, do, else, case, break},
  keywordstyle=\color{blue}\bfseries,
  ndkeywords={class, export, boolean, throw, implements, import, this},
  ndkeywordstyle=\color{darkgray}\bfseries,
  identifierstyle=\color{black},
  sensitive=false,
  comment=[l]{//},
  morecomment=[s]{/*}{*/},
  commentstyle=\color{purple}\ttfamily,
  stringstyle=\color{red}\ttfamily,
  morestring=[b]',
  morestring=[b]"
}

\DeclareFloatingEnvironment[
  fileext=lob,
  listname={List of Boxes},
  name={Code Listing},
  placement=htp,
]{BOX}

\mdfdefinestyle{mystyle}{
	hidealllines=true,
	leftline=true,
	innerleftmargin=10pt,
	innerrightmargin=10pt,
	innertopmargin=0pt,
}

\newmdenv[
topline=false,
bottomline=false,
skipabove=\topsep,
skipbelow=\topsep
]{siderules}

\newmdtheoremenv[style=mystyle]{frm-def}{Definition}

\crefname{frm-def}{definition}{definitions}

\newcommand{\shepherd}[1]{#1}
\def\name{Hacky Racers}

\def\rg{\textit{racing gadgets}}
\def\mg{\textit{magnifier gadgets}}

\def\rgsing{\textit{racing gadget}}

\def\mgsing{\textit{magnifier gadget}}
\title{\name{}: Exploiting Instruction-Level Parallelism to Generate Stealthy Fine-Grained Timers}

\author{Haocheng Xiao, Sam Ainsworth \\  University of Edinburgh \\  \textit{\{haocheng.xiao, sam.ainsworth\}@ed.ac.uk}
}

\date{}   

\begin{document}
\maketitle
\begin{abstract}

Side-channel attacks pose serious threats to many security models, especially sandbox-based browsers. While transient-execution side channels in out-of-order processors have previously been blamed for vulnerabilities such as Spectre and Meltdown, we show that in fact, the capability of out-of-order execution \emph{itself} to cause mayhem is far more general.

We develop \name{}, a new type of timing gadget that uses instruction-level parallelism, another key feature of out-of-order execution, to measure arbitrary fine-grained timing differences, even in the presence of highly restricted JavaScript sandbox environments. While such environments try to mitigate timing side channels by reducing timer precision and removing language features such as \textit{SharedArrayBuffer} that can be used to indirectly generate timers via thread-level parallelism, no such restrictions can be designed to limit \name{}. We also design versions of \name{} that require no misspeculation whatsoever, demonstrating that transient execution is not the only threat to security from modern microarchitectural performance optimization.

\shepherd{We use Hacky Racers to construct novel \textit{backwards-in-time} Spectre gadgets, which break many hardware countermeasures in the literature by leaking secrets before misspeculation is discovered. We also use them to generate the first known last-level cache eviction set generator in JavaScript that does not require \textit{SharedArrayBuffer} support.}

\end{abstract}

\section{Introduction}
The disclosures of Spectre~\cite{spectre} and Meltdown~\cite{meltdown} in 2018 have moved side channels into the mainstream. Various performance-optimization techniques in today's processors, such as transient execution, and caches~\cite{spectre, meltdown, spook.js, machineclear, spechammer, foreshadow, microscope, specreturns, fowardintime, interference, microop, ret2spec, netspectre,smotherspectre, fallout, zombieload, ridl, safecracker, streamline, cachetemplate, lruleaking, bufferoverflow, cachetelepathy, attackdir, prefetchattack, amdprefetchattack, sharedattackaes, doubletrouble}, have been shown to be exploitable by attackers to leak or exfiltrate information.
 
This has particularly manifested in browser security, where large amounts of untrusted sandboxed JavaScript code is executed. Although JavaScript is sandboxed, it is still vulnerable to most architectural attacks, e.g., Spectre~\cite{spectre, spook.js}, Rowhammer~\cite{rowhammerjs}, \emph{prime+probe}~\cite{spyinsandbox} and many others~\cite{clockticking, pixelperfecttiming, robustfingerprinting, aslrontheline, sharedattackaes, primescope}. Still, timing side-channel attacks rely on being able to reliably measure timing differences, and so the precision of the timers provided by browsers' APIs has been decreasing~\cite{firefoxtimer, chrometimer}. Although the level of assurance this mitigation actually gives has been hotly debated~\cite{performance.now,  primeprobe1, trustedbrowser, heretostay}, vendors still tend to resort to this whenever other more effective mitigations, such as patching processor microarchitecture, disabling hardware optimization techniques, or process-level isolation~\cite{siteisolation} are unavailable and/or degrade performance to too high a degree~\cite{cleanupspec, nda, DoM, invisispecbug, safespec, sdo}. 

A prime example of this is that when \textit{SharedArrayBuffer}, a JavaScript multi-threaded language feature, was found to provide an indirect fine-grained timer through counters and thread-level parallelism~\cite{fantastictimers}, it was temporarily removed from mainstream browsers~\cite{sharedarraybuffer} as a response to Spectre~\cite{spectre}. It is still blocked in recent mitigation proposals such as Chrome Zero~\cite{javascriptzero}, in the Tor Browser~\cite{Tortimer}, and in Firefox and Chrome for websites that have not opted in to cross-site isolation~\cite{chromesharedarraybuffer}.

In this paper we break the illusion of security brought by this timer-coarsening mitigation, and by the removal of \textit{SharedArrayBuffer}. We develop \name{}, a new class of timing gadgets that can be used to measure arbitrary fine-grained timing differences on out-of-order processors without any previously removed or potentially removable language features~\cite{sharedarraybuffer}, cross-thread contention~\cite{fantastictimers}, or indeed anything other than simple arithmetic operations, branches, loads, and coarse-grained timers.

While attacks such as Spectre~\cite{spectre} depend on the transient-execution capability of an out-of-order processor, and \textit{SharedArrayBuffer} timers on thread-level parallelism that can be disabled, we instead attack instruction-level parallelism (ILP): the ability for two or more data-independent instruction sequences from the \textit{same thread} to execute simultaneously, creating a race for which executes first.  
The key insight is that an attacker can use ILP to generate sequences of instructions that can be comparatively timed against each other, even in the absence of any direct source of time. The actual execution order between two racing instructions can be converted into long-lasting cache-state changes, such as an L1 eviction or reordering two cache fills, by \textit{racing}\shepherd{ (section \ref{sec: race})} two different sections of independent code against each other. This can then be \textit{magnified} \shepherd{ (section \ref{sec:magnifiers})} by repeatedly sampling the cache~\cite{treelru} or causing contention in the pipeline, to convert the fine-grained timing difference into a coarse-grained timing difference, undiminished by low resolution or noise.

This paper makes the following contributions:
\begin{enumerate}
  \item We propose a new method of exploiting out-of-order execution, \name, which use instruction-level parallelism to generate fine-grained timers.
  \item We introduce \rg{}, used to differentially time one event relative to another\shepherd{, and leave a state accordingly as an input for magnifier gadgets}.
  \item We introduce \mg{}
  , to amplify the timing difference between different micro-architecture states caused by a small difference, such as a speculative memory access, or the order of two cache fills.
  \item We prove the efficacy of \name{} through implementing several attacks in a browser without using \textit{SharedArrayBuffer}, including a new variant of Spectre V1~\cite{spectre}, SpectreBack, that can leak secrets backwards-in-time,
  to before any misspeculation is discovered.
  \item We demonstrate that \name{} can resurrect side-channel attacks thought to have been purged via timer coarsening, by implementing the first known eviction-set generator in JavaScript without \textit{SharedArrayBuffer}.
\end{enumerate}

\shepherd{To be clear, Hacky Racers are not as big a threat as transient-execution attacks such as Spectre~\cite{spectre}, as while transient-execution attacks directly leak secrets, Hacky Racers (and instruction-level parallelism in general) instead leak time. Still, that means that Hacky Racers can form the critical part in making information-leakage attacks (such as Spectre) feasible, by making information-recovery of them practical even in extremely restricted environments.}

\noindent We have disclosed our findings to the Tor Browser, Chrome, Firefox, and Cloudflare. The Hacky-Racers attack code is open-sourced at \url{https://github.com/FxPiGaAo/Hacky-Racer}.
\section{Background and Related Work}

Here we first introduce how timers are used to facilitate side-channel attacks. We then discuss current methods of timing in browsers and their mitigations, followed by background on the out-of-order execution of modern processors that we use to generate the instruction races used in \name{}. Finally, we discuss Spectre attacks.

\subsection{Timing in JavaScript Attacks}\label{src:jstiming}
Timers are typically used in JavaScript attacks in two places: in the receiving stage of a side channel (e.g. to time the presence/absence of a cache line~\cite{spook.js} or execution-unit contention~\cite{spectrerewind}), in the last-level cache (LLC) eviction set (EV) profile stage~\cite{primescope, dynamicallyfindev, theoryfindsets, rowhammerjs, spyinsandbox, practical, doubletrouble, leakyway} (to allow efficient preparation of the cache before an attack), or both. 

The precision of the timer matters when the attacker needs to time the access latency to one or multiple addresses to distinguish whether a cache miss exists or not. Last-level-cache (LLC)-based channels~\cite{javascriptzero, spyinsandbox, cachetemplate}, such as \emph{Flush+Reload}~\cite{flushreload}, \emph{Prime+Probe}~\cite{primeprobe2005,primeprobe2006, llcpractical,primeprobe1}, \emph{Evict+Time}~\cite{primeprobe2006}, \emph{Evict+Reload}~\cite{cachetemplate}, \emph{Reload+Refresh}~\cite{reloadrefresh, lruleaking} and \emph{Evict+Prefetch}~\cite{prefetchattack}, are prevalent, since they are still effective even if the attacker and victim never share the same core. Though the timing of an LLC miss, which is used in the eviction-set profiling and the secret transmission stage, is more coarse grained than an L1 cache miss, its timing difference is still around 100ns, and so cannot be timed natively with the coarse-grained timers (\cref{sec: jstimer}) in today's browsers.

A precise timer is critical to some other (non-timing-side-channel) attacks, such as rowhammer.js~\cite{rowhammerjs} and Spook.js~\cite{spook.js}, as they also require LLC eviction-set profiling to actively trigger victim cache misses from the attacker's side.
Finally, there are other side-channel attacks that require timing information such as
website fingerprinting~\cite{primeprobe1, automatedfingerprint, robustfingerprinting, spyinsandbox}, inferring neural-network architectures~\cite{cachetelepathy} and breaking AES~\cite{sharedattackaes}. 

\subsection{Timers in Browsers and Mitigation}\label{sec: jstimer}
Sources of time in browsers can be generated intentionally by APIs, or unintentionally by other system behavior~\cite{fantastictimers, soklosttime}.

\textbf{Performance.now()}  is the most precise timer provided natively by browsers' APIs. Its precision has changed over time as a response to attacks. For example, its resolution was decreased to 5$\mu$s in both Chrome and Firefox in 2015 as the first browser side-channel attack~\cite{spyinsandbox} emerged. In early 2018, Spectre~\cite{spectre} forced vendors to further decrease its precision to 2ms in Firefox 59, and 100ms in Chrome with a 100ms jitter~\cite{soklosttime}. After site isolation~\cite{siteisolation}, a mitigation to remove secrets from the address space, was added in Chrome and Firefox, the precision restriction was loosened to 5$\mu$s plus jitter in Chrome~\cite{chrometimer} and 5$\mu$s in Firefox~\cite{firefoxresolution} within the same origin\footnote{Resolution can also be customized to a larger value by enabling privacy.resistFingerprinting~\cite{firefoxtimer} in Firefox, and 100$\mu s$ (from the default 5$\mu s$) in non-isolated contexts in Chrome.}.

\textbf{SharedArrayBuffer}  was introduced by ECMAScript 2017~\cite{ecma2017} to facilitate cross-thread communication.
It provides shared memory between multiple workers, enabling simultaneous data reading and writing between threads. Based on this feature, Schwarz et. al. ~\cite{fantastictimers} built a high-resolution timer with precision around 2 - 15 nanoseconds. A shared value among a main thread and a subthread is allocated in this \textit{SharedArrayBuffer}. The subthread serves as a counting thread, incrementing the value continuously in an infinite loop.
The main thread then gets timestamps by reading the value.

Similar to the high-resolution performance.now() API, the \textit{SharedArrayBuffer} feature was also disabled~\cite{disablesab} as a response to Spectre~\cite{spectre} and re-enabled after the implementation of site isolation in Chrome and Firefox only for cross-origin isolated pages~\cite{chromesharedarraybuffer}. It is still disabled in Tor~\cite{Torsharedarraybuffer} and Chrome Zero~\cite{javascriptzero}.

\textbf{Others}
Apart from lowering precision, other countermeasures have been proposed, such as fuzzy timers~\cite{eliminating, trustedbrowser} and limiting shared memory and message passing. In fuzzy time, the clock edge is randomly perturbed, further reducing the final observed precision.
This was thought to mitigate the edge-thresholding technique~\cite{fantastictimers}, which otherwise has the potential to recover the resolution by adding extra instructions to reach the threshold.

\subsection{Repetition and Magnifier Gadgets}
To observe fine-grained timing difference with a coarse-grained timer, the signal must in some way be amplified. Here we divide such methods into two categories: \textit{repetition gadgets}, which repeat the attack many times, and \textit{magnifier gadgets}, which require the attack to be performed only once.

Repetition gadgets repeat the whole extract-then-transmit process to accumulate the timing difference from each iteration. For example, Skarlatos et al.~\cite{microscope} denoise their SGX attack by repeatedly triggering replay on illegal instructions without actually incurring a page fault. McIlroy et al.~\cite{heretostay} propose a gadget to create timing difference for Spectre gadgets~\cite{spectre} that can be observed by the coarse-grained timer in a browser, by repeatedly triggering the attack. \shepherd{Schwarzl et al.}~\cite{dynamicprocessisolation} \shepherd{implement a similar attack on Cloudflare Workers~\cite{cloudflareworkers}, the serverless execution framework, where timing can only be achieved by a remote timing server.} 
However, repetition on its own has the risk of cancelling out the timing difference between different stages (\cref{sub: naive repetition}).

In contrast, a magnifier gadget first extracts and transmits the secret into state differences once, and then executes instructions that both maintain the state difference and accumulate the timing difference at the same time. Therefore, this gadget will not suffer from noise from stages such as branch mistraining. We repurpose the magnifier gadget introduced by the leaky.page Spectre attack~\cite{treelru}, which attacks PLRU caches, to time arbitrary sequences of instructions, and introduce new magnifier gadgets that avoid any transient execution, and work for other cache replacement strategies (\cref{sec:magnifiers}).

Orthogonally, attackers can also leave multiple state differences~\cite{spring}
after the secret is extracted once, which can also increase the timing difference by a constant (typically limited) multiple -- though further timer coarsening by a similar constant can mitigate the effects.

\subsection{Out-of-Order Execution}\label{sec: outoforder}
Out-of-order execution is an almost ubiquitous~\cite{spectre} processor microarchitectural optimization technique, designed to exploit instruction-level parallelism, by executing many independent instructions from the same thread simultaneously when there are no data dependencies between them. This optimization must obey the programmer's sequential view of execution, and so instructions commit in-order at the backend of the pipeline. Still, the true out-of-order nature of execution may still leave traces in structures outside the programmer's model, such as the cache, which we can pick up to infer which executed first.

In this paper, we define a \emph{path} (\cref{sec: path}) as an instruction sequence that currently does not have any external data dependence, and could thus be processed independently of any instructions outside of that sequence. This independence allows execution to flow within separate paths simultaneously.

\subsection{Speculative Execution Attacks}

To find useful work, an out-of-order processor must speculate on the future flow of instructions, such as via branch predictors. This incorrect execution will be rolled back, but may still leave traces that can be found by an attacker. 

Misspeculated (or transient) execution was famously attacked by Spectre~\cite{spectre}: the branch predictor or branch target buffer can be poisoned to cause incorrect instructions (such as out-of-bounds accesses to arbitrary parts of the address space) to execute speculatively, and leave traces in the cache. Meltdown~\cite{meltdown} similarly attacked the behavior of some Intel cores to incorrectly forward on the values of exceptional loads to other transient instructions.

\name{} are different. Though transient execution makes the construction of its gadgets simpler, the attack requires no transient execution at all. The \emph{correct} execution of instructions\footnote{Concurrent to this work, Rokicki et al.~\cite{portcontetionILP} also exploited ILP to build a port-contention channel in JavaScript that can be used to determine a user's CPU model. Their attack can be used to replace the simultaneous multi-threading (SMT) feature relied by the original attack with ILP contention, but still uses a \textit{SharedArrayBuffer} based timer, which Hacky Racers can replace.}, executed in the wrong order, is still a threat to browser security. 
\name{} also provide new perspectives on transient execution attacks. In section \ref{sec: JS backward-in-time}, we will show new racing gadgets that allow leaking of data in ways that many proposed mitigations cannot eliminate. Many other speculative execution attacks based on contention, of either arithmetic logic units (ALUs) in the core~\cite{spectrerewind} or miss status holding registers (MSHRs) in the cache~\cite{interference}, could also be validly formulated as \name{} racing gadgets.

\section{Threat Model}
We assume an arbitrary side-channel attack in a JavaScript application, such as an advertisement in a web browser trying to access secrets outside its sandbox, that needs to time one or several operations in order to succeed, such as the probe operation in a prime+probe~\cite{primeprobe2005}. We further assume that the timing difference brought about by this side channel is fine-grained -- of the order of 100 nanoseconds or smaller, thus cannot be observed by current browsers' JavaScript native timers. We also assume the attacker can execute any valid JavaScript code within their sandbox, but that any timers or other forms of time (\cref{sec: jstimer}), are limited to $5\mu s$ granularity. The majority of our attacks also work for as coarse a granularity as has ever been introduced in a browser (100ms) and higher, but we use $5\mu s$ as the threshold for success of an attack. We do not require the attacker to have access to any cross-thread primitives. The attacker will succeed if they are able to measure this difference and thus construct an information channel. This gives \name{} the potential to resurrect or accelerate attacks that were previously hindered by timing obfuscation~\cite{trustedbrowser, javascriptzero}.

We assume the attacker is executing on a processor with out-of-order execution, as is typical on modern devices~\cite{spectre}, able to execute tens-to-hundreds of instructions into the future. The instruction length of operations to be measured is limited by the reorder buffer of a processor in practice. However, we find that very few concurrently executing instructions are required for success.

\shepherd{Our threat model could also be expanded to other environments where a fine-grained timer is unavailable, such as Cloudflare Workers~\cite{cloudflareworkers} and, Apple M1 processors and other ARM systems~\cite{pacman, armageddon}. We leave these to future work for simplicity.} 
\section{Path Construction}\label{sec: path}

\begin{figure}
    \centering
    \includegraphics[width=0.45\textwidth]{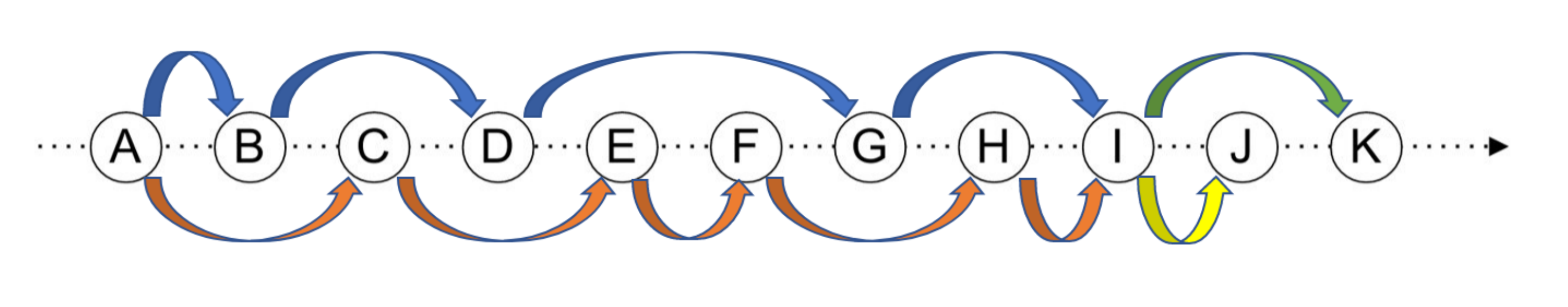}
    \caption{Paths within instructions. Instructions are placed in program order on this dotted timeline. Data-dependence is indicated as arrows, and is colored in group of paths. When instruction A's result is ready, two independent \emph{paths} could execute in any order: \{B, D, G\} and \{C, E, F, H\}.}
    \label{fig: instruction path}
\end{figure}

Here we define \textit{paths} as a unit of simultaneous execution \textit{within} a single thread (exemplified in \cref{fig: instruction path}) on an out-of-order processor. \shepherd{Between two independent paths $path_a$ and $path_b$, there must be no data dependencies, and thus they may execute entirely in parallel with each other on an out-of-order processor. Likewise, no instruction within the path should be data dependent on any instruction outside the path, except for instructions that come before the entire path in program order. Within a path, there may be one or more \textit{chains} which are connected via data dependence, such that no two instructions within each \textit{chain} can execute simultaneously or out-of-order. Note that \textit{paths} are defined in order to simplify and reason the construction of our Hacky Racer, and there could be multiple combinations to separate one instruction sequence into \textit{paths} while only one of them is meaningful.}

\shepherd{We first introduce the construction process of \textit{paths} that satisfy the synchronization and concurrency requirements of Hacky Racers.} {We then demonstrate how to embed the expression whose timing we would like to observe, named the \textit{target expression} or ${Expr_t}$, into a path as the first step of our Hacky Racer.}

\shepherd{Ultimately, this path will compete against another path, the \textit{baseline path}, as part of the \textit{racing gadgets} of section \ref{sec: race}, which transform the race result of the two paths into a state change in the system. This state change can then be input into magnifier gadgets to amplify the side channel to something observable via coarse-grained timer, even after only a single attack instance.}

\subsection{Path Synchronization}\label{sec: concurrency}

Code listing~\ref{box:a} shows an example of a construction of two independent paths, $path_a$ and $path_b$, that are eligible for execution simultaneously. In this example, each path consists of a single chain.
\shepherd{To synchronize their starting execution, we ensure the first instruction in the chains in both paths have a common data dependency, in instruction A\footnote{\shepherd{Where paths have multiple chains within them as in figure \ref{fig: expression path}, the first instruction in every chain must depend (transitively) on this common dependency.}}. We ensure that instruction A,} loading \textit{Array[0]}, will incur \shepherd{a cache miss that delays executions of both paths to avoid} frontend fetch/decode contention, meaning all instructions will have reached the out-of-order backend by the time any is eligible for execution.  As a result, $path_a$ and $path_b$ start at the same time and run in parallel.

\begin{BOX}[t]
\centering
\begin{lstlisting}
function myFunction() {
  var A = array[0];
  var B = arrayA[A]; //path A 
  var C = arrayB[A]; //path B 
  var D = arrayA[B]; //path A 
  var E = arrayB[C]; //path B 
  var F = arrayA[D]; //path A 
  var G = arrayB[E]; //path B 
  var H = arrayA[F]; //path A 
  var I = arrayB[G]; //path B 
  var J = H + I;
  return J;
}
\end{lstlisting}
\caption{An example for two synchronized paths. These form two independent chains of operations that will run concurrently on an out-of-order processor: knowledge of which finishes first can be used as a side channel.\label{box:a}}
\end{BOX}

\subsection{Expression Embedding}
In addition to synchronizing the start of each path's execution, to observe the execution timing of the \textit{target expression}, it should be embedded into a single path, named the \textit{measurement path}. The path may also be required to end with an attacker-defined result \shepherd{as its \textit{terminator instruction}, transitively dependent on the final instruction of every \textit{chain} within the path. In practice this terminator instruction will serve either as a branch-condition variable or as a data-access index, to be used as a side channel to infer their relative timing, as will be introduced in section \ref{sec: race}.} To satisfy these two restrictions, we construct both a \textit{pre-extension} and \textit{post-extension} around \shepherd{every chain within} the original \textit{target expression}. The \textit{pre-extension}'s output encodes the input data for the \textit{target expression}, ensuring that all data inputs to the function depend on a single \textit{head instruction} at the start of the path, while the \textit{post-extension} converts the original outputs of the \textit{target expression} into inputs to an attacker-controlled instruction, to ensure that this instruction only executes following the completed execution of all outputs.

\begin{figure}[t]
    \centering
    \includegraphics[width=0.45\textwidth]{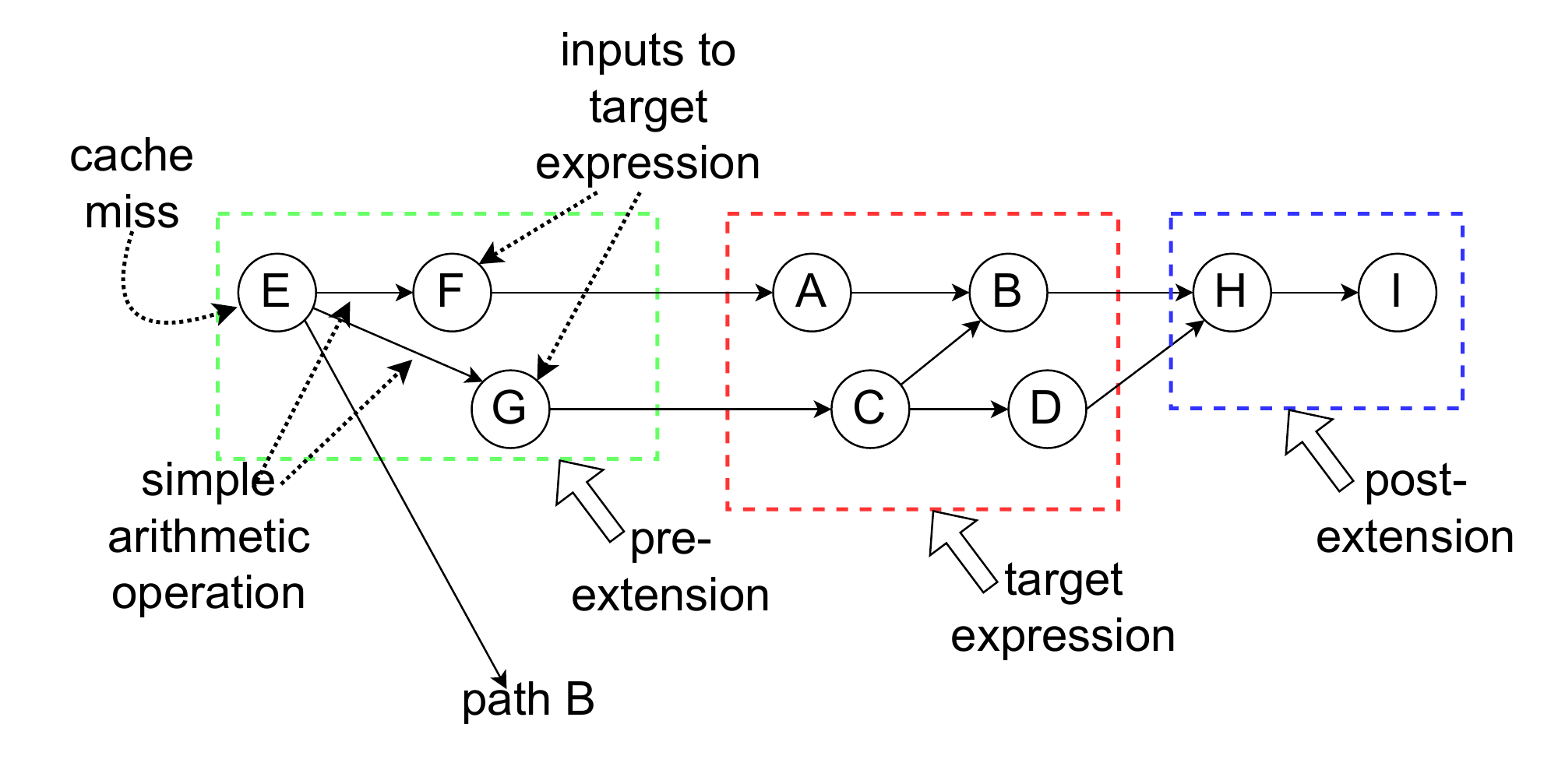}
    \caption{The \textit{target expression} to be measured is embedded into a path. \textit{Pre-extension} and \textit{post-extension} instructions are shown in the green and blue box, wrapping the target expression. Instruction E, the beginning of the path, will incur a cache miss, and be depended by $path_b$.}
    \label{fig: expression path}
\end{figure}

In our example, The \textit{target expression} is shown both in the red-dotted box of figure \ref{fig: expression path} and in the \textit{OriginFunc()} of code listing \ref{box:b}. The transformed function contains the \textit{measurement path} transformed from the \textit{target expression}. Between the \textit{pre-extension} and \textit{target expression}, the indices used to load A and C in the \textit{OriginFunc()} are replaced by F and G, which encode each index as a data dependency of E. As is discussed in section \ref{sec: concurrency}, the load of E will suffer a cache miss to synchronize the starts of the two paths. In the post-extension, var H is created to depend on both B and D, ensuring the completion of the whole path only after B and D have finished execution. Finally, in var I, an offset is added to H to achieve the desired output (for example, the access of a specific cache line, or branch condition, to create the side channel that indicates completion). \shepherd{Thus, the \textit{target operation} is embedded into a path that produces a particular output only once the full operation has completed. In the next section, we will describe how to synchronize it against a competing path, by using this chosen output as a side channel to infer ordering between the two paths.}

\begin{BOX}[t]
\centering
\begin{lstlisting}
function OriginFunc(x, y) {
  var A = array[x];
  var C = array[y];
  var B = array[A+C];
  var D = array[D];
  return B + D;
}

//construction of path A
function TransformedFunc(x, y) {
//pre-extension
  var E = array[0]; //E = array[0] = 0
  var F = E + x; //F = x (input 1)
  var G = E + y; //G = y (input 2)

//modified target expression
  var A = array[F];
  var C = array[G];
  var B = array[A+C];
  var D = array[D];
  
//post-extension
  var H = B + D; //depends on both outputs
  var I = H + delta; //I = attacker-predefined
  
//path B is omitted here, which also depends on E
  ...
  return;
}
\end{lstlisting}
\caption{The corresponding code example for fig. \ref{fig: expression path}, where the \textbf{target expression} is embedded into a path. \emph{OriginFunc()} is the target expression, while the \emph{TransformedFunc()} is the transformed version, with \textit{pre-extension} to cause both Paths in OriginFunc to run simultaneously, and post-extension to leave a microarchitectural trace that can be picked up as the unique end of the target expression. \label{box:b}}
\end{BOX}

\section{Racing Gadgets}\label{sec: race}
We develop \rg\ to measure timing \textit{relatively} with reference to a \emph{path} with a known constant execution time, which we call a \emph{baseline path}, $path_b$. The target expression to be measured, $Expr_t$, is placed in the \emph{measurement path}, $path_m$, starting at the same time as the \emph{baseline path}.
Thus, the \emph{path} with a shorter execution time will finish first. This \emph{completion order} between these two \textit{paths} leaves a micro-architectural difference that will serve as the input of the \textit{magnifier gadget}.

\subsection{Transient Presence/Absence (P/A) Racing Gadget}\label{sec: p/a racing gadgets}
This gadget converts the \emph{completion order} into whether there is a transient cache access to a specific address or not. This is achieved by altering how quickly a mispredicted branch is corrected, depending on the operation time that we are interested in. The cache-state change caused by the transient access serves as input to the corresponding \textit{magnifier gadget}.

Assume $T$ is threshold which we would like to distinguish whether the execution time of $Expr_t$ is larger or smaller than. We prepare two operation sequences $path_m(Expr_t, x)$ and $path_b()$\footnote{The construction to satisfy these restrictions is discussed in section \ref{sec: path}.}. The variable $x$ \shepherd{ controls the final data output of} $path_m$.
$path_m(Expr_t, x)$ satisfies the following requirements:

\noindent \fbox{\parbox{\dimexpr\linewidth-2\fboxsep-2\fboxrule\relax}{a) $path_m(Expr_t, 0) = 1$ and $path_m(Expr_t, 1) = 0$. \\ b) previous execution of $path_m(Expr_t, 0)$ will not affect the execution time of $path_m(Expr_t, 1)$.
}}
Requirements a) and b) are needed due to the training phase, since its output serves as a condition variable to trick the branch predictor into executing transient code. We set x = 0 during the training phase and x = 1 during the detection phase.

\noindent \fbox{\parbox{\dimexpr\linewidth-2\fboxsep-2\fboxrule\relax}{c) $Time(Expr_t)_{low} < T <  Time(Expr_t)_{high} \Rightarrow $ \scalebox{0.94}{$Time(path_m(Expr_t, 1))_{low} < T' < Time(path_m(Expr_t, 1))_{high}$}
}}
Requirement c) converts the  execution-time comparison between $Expr_t$ and T into one between $path_m(Expr_t, 1)$ and T', where T' accounts for the extra time taken in \textit{pre-extension} and \textit{post-extension}. $Path_b$() is generated so that its execution time is constant and equal to T', so that the race can distinguish whether $Time(Expr_t)$ is \textit{high} or \textit{low}.

\noindent \fbox{\parbox{\dimexpr\linewidth-2\fboxsep-2\fboxrule\relax}{
d) No instruction in $path_b()$ can have a data dependency on any instruction in $path_m(Expr_t, 1)$\shepherd{, and vice versa}\\
e) $path_b()$ and $path_m(Expr_t, 1)$ should be started at almost the same time.

}}
Requirement d) is implied by the definition of \emph{path}, and is necessary to allow the two paths to execute independently (and thus race against each other), while requirement e), to make the race between the two fair (and thus increase precision), is satisfied by letting the first instruction in both paths to be dependent on the same cache miss (section \ref{sec: concurrency}).
 
Finally we construct the timing gadget as follows ($\longmapsto$ indicates that access[A] has a data dependence on the result of $path_b()$, thus it will be executed after all of the instructions in $path_b()$ have executed even on an out-of-order processor):

\noindent \fbox{\parbox{\dimexpr\linewidth-2\fboxsep-2\fboxrule\relax}{
\setlength{\parindent}{5ex}\noindent if($path_m(x)$) \par $path_b()\longmapsto access[A];$}}%

Before this gadget is executed, we train the branch predictor through executing this code snippet with x = 0. When we execute it with x = 1, the mispredicted $path_b()$ will be executed in parallel with $path_m()$ until one finishes. If $Time(Expr_t)>T$, the memory request of access[A] will be sent after $path_b()$ finishes before $path_m()$. Otherwise when $path_m()$ finishes first, it will roll back before it reaches $access[A]$.

\subsection{Non-transient Reorder Racing Gadget}\label{sec: non-transient rg}

Here we introduce another \rgsing{} that does not rely on transient execution, unlike Spectre~\cite{spectre}. No misspeculation is required for this gadget, so it is unaffected by hardware mitigation schemes for Spectre~\cite{ghostminion,invisispec,stt} and demonstrates that we are attacking a fundamentally different property of out-of-order execution. Here, inspired by a gadget used as part of speculative interference attacks~\cite{interference} to defeat invisible transient caching mechanisms for Spectre resistance~\cite{invisispec}, we convert \emph{completion order} into the \textit{relative order} of two memory accesses, A and B. We construct the gadget as follows:

\noindent \fbox{\parbox{\dimexpr\linewidth-2\fboxsep-2\fboxrule\relax}{
\setlength{\parindent}{5ex}\noindent $path_m()\longmapsto access[A];$ \par \noindent $path_b()\longmapsto access[B];$}}%

And with the following requirements:

\noindent \fbox{\parbox{\dimexpr\linewidth-2\fboxsep-2\fboxrule\relax}{a) $Time(Expr_t)_{low} < T <  Time(Expr_t)_{high} \Rightarrow $ \scalebox{0.94}{$Time(path_m(Expr_t))_{low} < T' < Time(path_m(Expr_t))_{high}$} \\ b) No instruction in $path_b()$ can have a data dependency on any instruction in $path_m()$\shepherd{, and vice versa} \\
c) $path_b()$ and $path_m()$ should be started at the same time.
}}
These are similar to those of the transient P/A racing gadget except that the first two restrictions are removed, as this gadget does not include branches. Because $path_b()$ and $path_m()$ start execution simultaneously, the one which completes first will issue the subsequent memory access earlier than the other.
\section{Magnifier Gadgets}
\label{sec:magnifiers}

The \textit{racing gadgets} of \Cref{sec: race} place the system in one of two different micro-architectural states: one to transmit a 1, the other to transmit a 0\footnote{More generally, this may be N different states.}. This state difference serves as the input of the \textit{magnifier gadget}. Our magnifier gadgets use this state as input to create a cascading time difference, typically (but not exclusively) by creating a large number of cache misses for one of these two states, and a large number of cache hits for the other. We first introduce two methods to compose a magnifier gadget in a (widely used~\cite{pseudolru}) tree-based pseudo-last-recently-used (PLRU) L1 cache -- one for the P/A transient racing gadget, and one for the reorder-based non-transient gadget. Finally, we illustrate magnifier gadgets on both a cache with arbitrary replacement policy, and one not using the cache at all -- showing that changing the replacement policy is no cure for \name{}, even if specific policies may make the gadgets faster and simpler to implement.

\shepherd{Note that our \textit{magnifier gadgets} simply transform a particular  \textit{state difference} to a large \textit{timing difference}; this is not limited to being from the output of the \textit{racing gadgets} from section \ref{sec: race}. Indeed, the first gadget is taken from a Spectre attack~\cite{treelru} in the literature, which generates the state difference directly. Rather, the \textit{racing gadgets} are used to convert \textit{arbitrary expressions} to a particular state-change format, as input to a given magnifier. This generic timing capability can manifest in subtle ways: for example in section \ref{ssec:llc}, where a racing gadget allows us to convert a last-level cache side channel into an L1-cache side channel, thus allowing us to use an L1-cache magnifier and thus demonstrate the first eviction-set generator in JavaScript without SharedArrayBuffer.}

\begin{figure}[ht!]
\subfloat[The initial cache state is shown in fig \ref{fig: plru}.1. Before A was accessed, B was the eviction candidate (EVC), and thus is replaced by A. The pseudo-LRU state is flipped down the accessed path, causing the EVC to switch to E (fig \ref{fig: plru}.2). In fig \ref{fig: plru}.3, since access C is a cache hit, nothing is evicted, however the EVC still changes, flipping the arrows down the accessed path between it and fig \ref{fig: plru}.4.]{%
  \includegraphics[width=9cm,trim={0 0 0 1.3cm},clip]{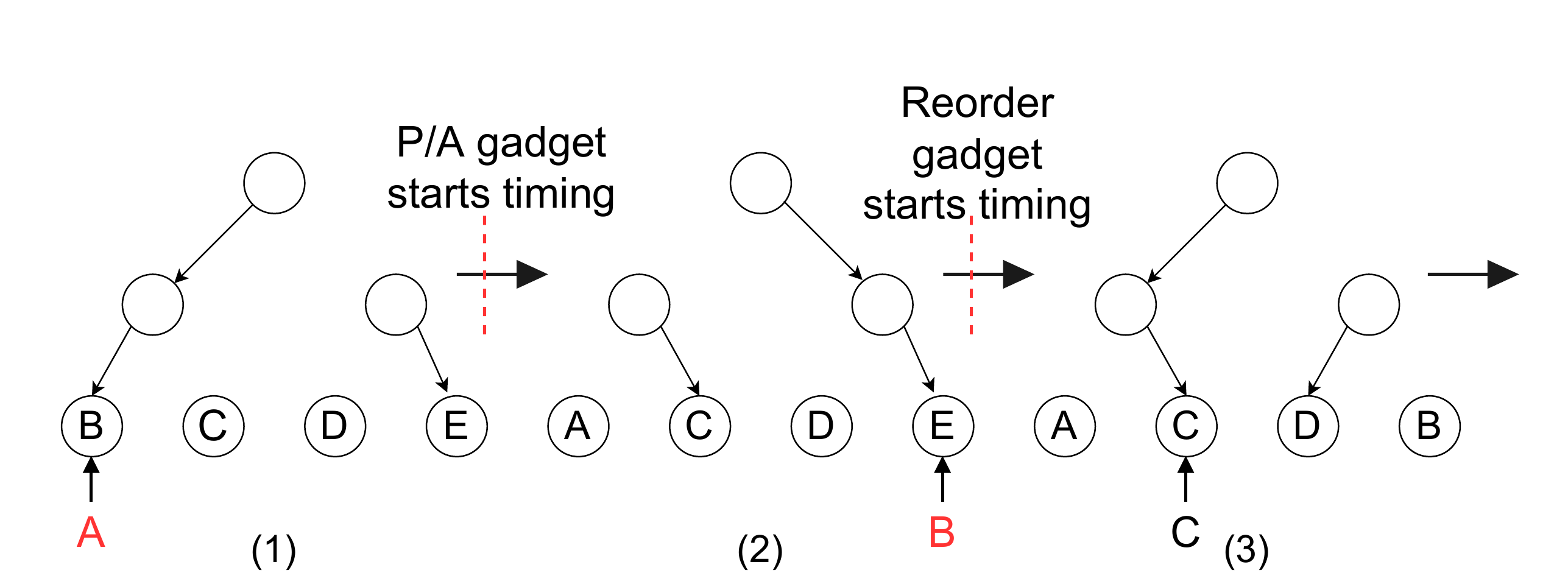}%
}

\subfloat[After accessing E in fig \ref{fig: plru}.4, A becomes the new EVC (fig \ref{fig: plru}.5). Since we want to keep A in the cache to allow repeated measurement, C is accessed next, making B the new EVC in fig (\ref{fig: plru}.6), causing D to evict B rather than A.]{%
  \includegraphics[width=9cm,trim={0 0 0 1.3cm},clip]{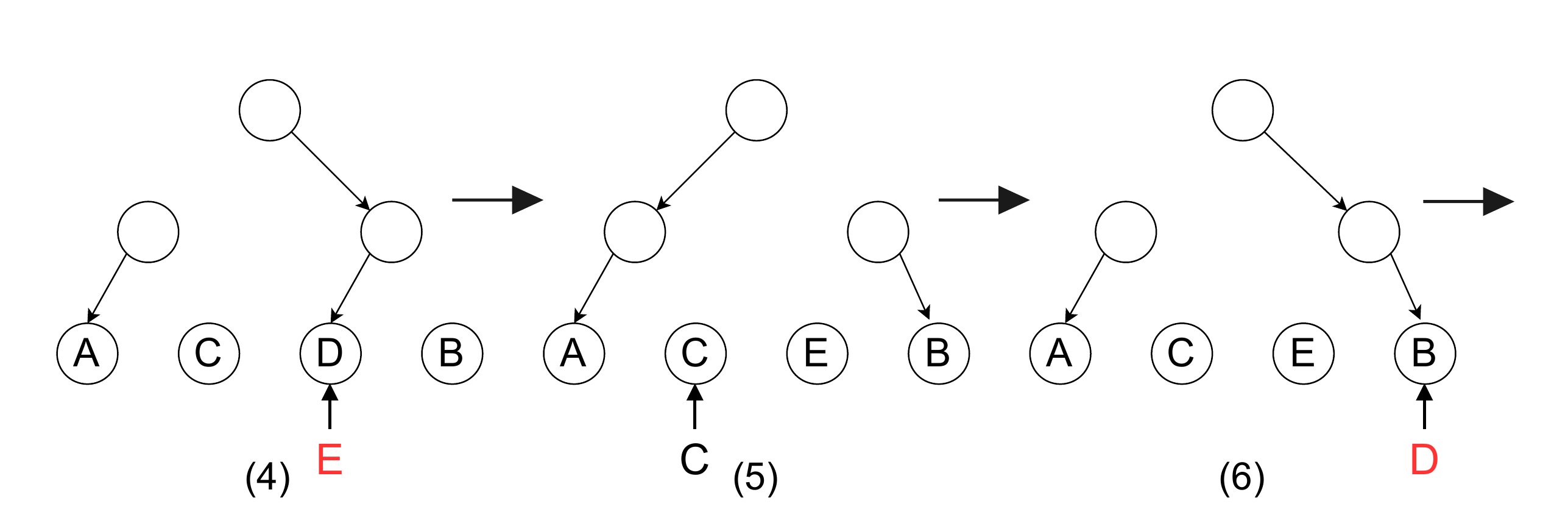}%
}

\subfloat[Since A becomes the EVC again in fig \ref{fig: plru}.7, C needs to be accessed again to flip the pseudo-LRU state at the top of the tree. Now both the cache state and the access pattern go back to that in fig \ref{fig: plru}.2, allowing the sequence to be repeated indefinitely without a new access to A.]{%
  \includegraphics[width=9cm,trim={0cm 1cm 0cm 1.3cm},clip]{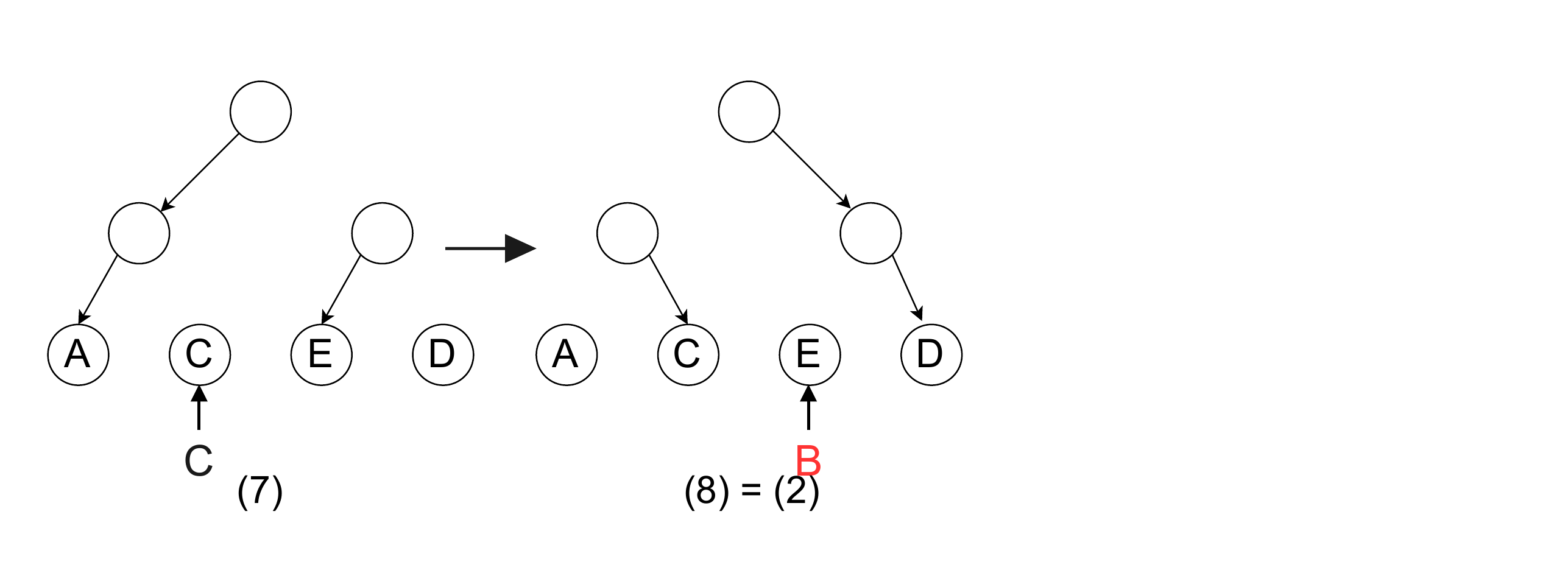}%
}

\caption{Cache replacement state and data changes caused by each access of the PLRU Gadget for both 1) \textit{Presence/Absence Input} (\cref{pa gadget}) when access A is present, and 2) \textit{Reorder Input} (\cref{plru reorder input}) when A is inserted before B. If A is in the cache in the former case, or inserted before B in the latter case, the accesses to B, D and E will repeatedly miss, and the access to C repeatedly hit, and yet the PLRU state prevents them from ever evicting A. The timing starts at the beginning of the corresponding magnifier gadget. Each leaf node represents a cache line in the set and data in it. The arrows within each sub-figure composes one path from root to the leaf, pointing to the eviction candidate. The access pattern and the cache states repeat in a period of 6 accesses, as is shown in (2) - (7).}
\label{fig: plru}
\end{figure}

\subsection{PLRU Gadget for Presence/Absence (P/A) Input} \label{pa gadget}

This first magnifier gadget is inspired by Röttger and Janc~\cite{treelru}, who used a similar strategy as part of the leaky.page proof-of-concept of a Spectre V1 JavaScript attack. We repurpose it to time arbitrary execution. It utilizes the tree-based Pseudo-Least-Recently-Used (PLRU)~\cite{pseudolru} cache replacement policy, prevalent on modern CPUs~\cite{treelru}, to amplify the timing difference of a single victim access. PLRU policies implement a binary tree structure to approximate a Least Recently Used (LRU) replacement policy. For a clear illustration, we take the PLRU cache's set associativity $W$ = 4 as an example, and assume all of the following accesses are mapped into the same cache set. As is shown in figure \ref{fig: plru}, each leaf node represents a cache line within that set, and the letter within that node indicates the data currently filled in that line. Cache misses are marked in red and those arrows from the root to the leaf always composes one path pointing to the eviction candidate. Every time an access happens to a specific location within that set, it will flip arrows on its path.

The idea behind this gadget is that if the target memory address is brought into the cache by the racing gadget, it will be kept in the L1 cache for the entire magnifier gadget, such that the observed capacity of the set (in timing terms) is less by one than the elements of the following access pattern. Here we construct the access pattern to be a repetition of four locations: (B, C, E, C, D, C). If A is absent from the cache, no miss will occur. If it is present, the PLRU gadget will never evict it, and thus the cache state and the access pattern will repeatedly go back to the previous state in figure \ref{fig: plru}.8, forming a cycle from figure \ref{fig: plru}.3 to \ref{fig: plru}.8. Therefore, cache misses happen every other access, as is shown in figures \ref{fig: plru}.2, \ref{fig: plru}.4, and \ref{fig: plru}.6.

\subsection{PLRU Gadget for Reorder Input}\label{plru reorder input}
This gadget takes advantage of the fact that A will be put in different locations within the cache set based on whether A or B is accessed first. As a result, A will always be kept in the cache if A arrives \textit{before} B, and yet will be evicted if A arrives \textit{after} B. Similar to the PLRU gadget for P/A input, the occupation of A will result in a series of cache misses, while the absence of A will present a series of hits, from a sequence of accesses not involving A.

\begin{figure}[t]
    \centering
    \includegraphics[width=0.45\textwidth]{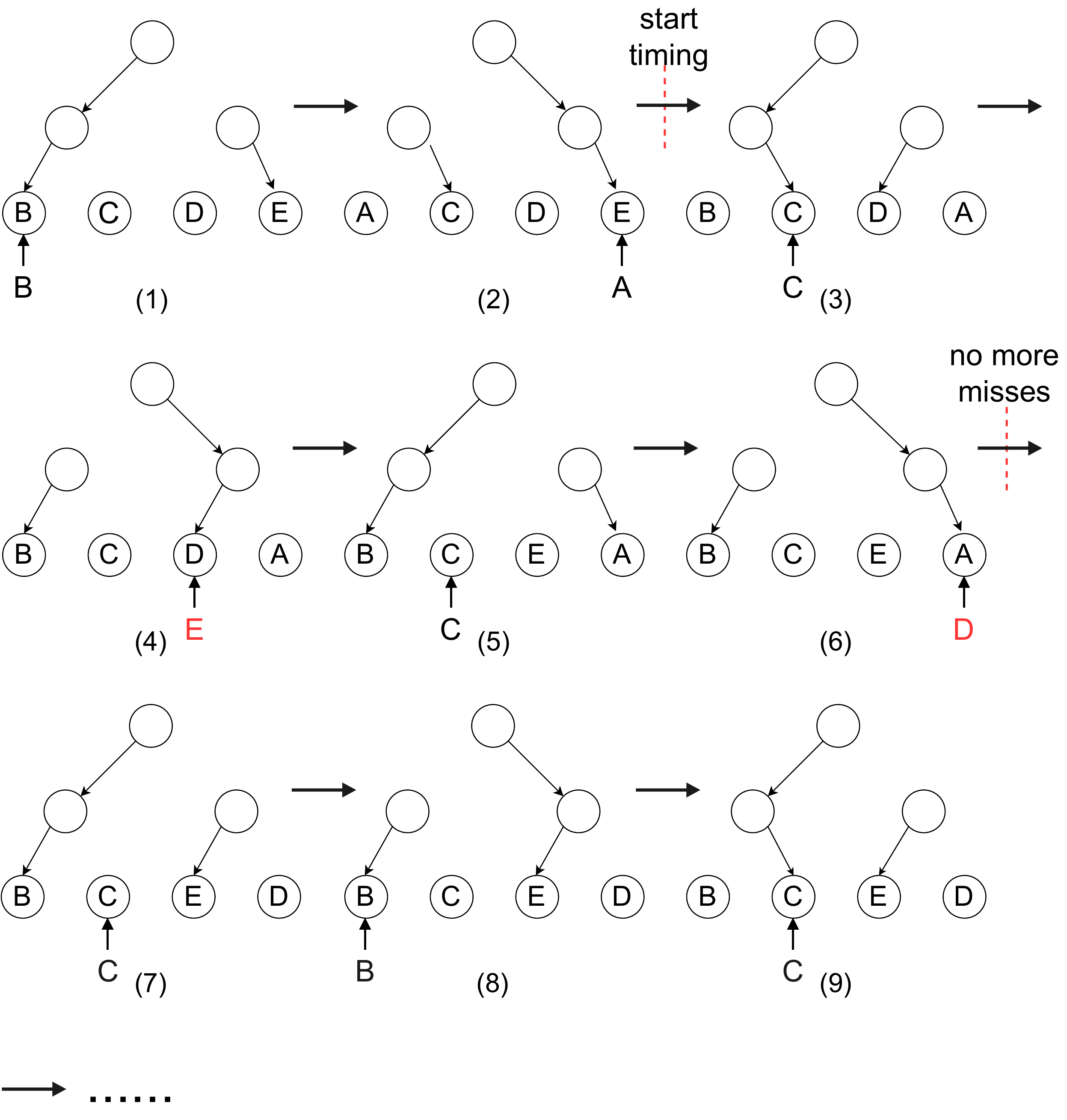}
    \caption{Cache state changes caused by each access of PLRU Gadget for P/A Input. Here B is inserted before A as is shown in the first two subfigures. A is evicted at (6), thus no more misses after that.}
    \label{fig: b_a}
\end{figure}

The following accesses pattern is a repetition of (C, E, C, D, C, B).  Figure \ref{fig: plru} and \ref{fig: b_a} contrast the cache states of two different relative orders between A and B during the execution of the magnifier gadget. Both start with the same initial state, but proceed differently depending on how A and B are reordered. When A is accessed before B, the state-change flow is the identical to that in the P/A PLRU gadget, despite requiring no transient execution and thus meaning the racing gadget accesses exactly the same cache lines regardless of transmitted signal. The difference is subtle: B in the Reorder Input magnifier gadget is a part of the racing gadget, whereas it is part of the magnifier gadget in the P/A version. 
By contrast, when B is accessed first, A will be evicted after several accesses, as is shown in \cref{fig: b_a}. After that, there will be no more cache misses, since all following accesses fit into the cache.

\begin{figure}[ht]
\captionsetup[subfloat]{farskip=-4pt,captionskip=-13pt,width=.99\columnwidth}
\subfloat[The timeline of two paths when no interference occurs.]{
        \includegraphics[width=\columnwidth]{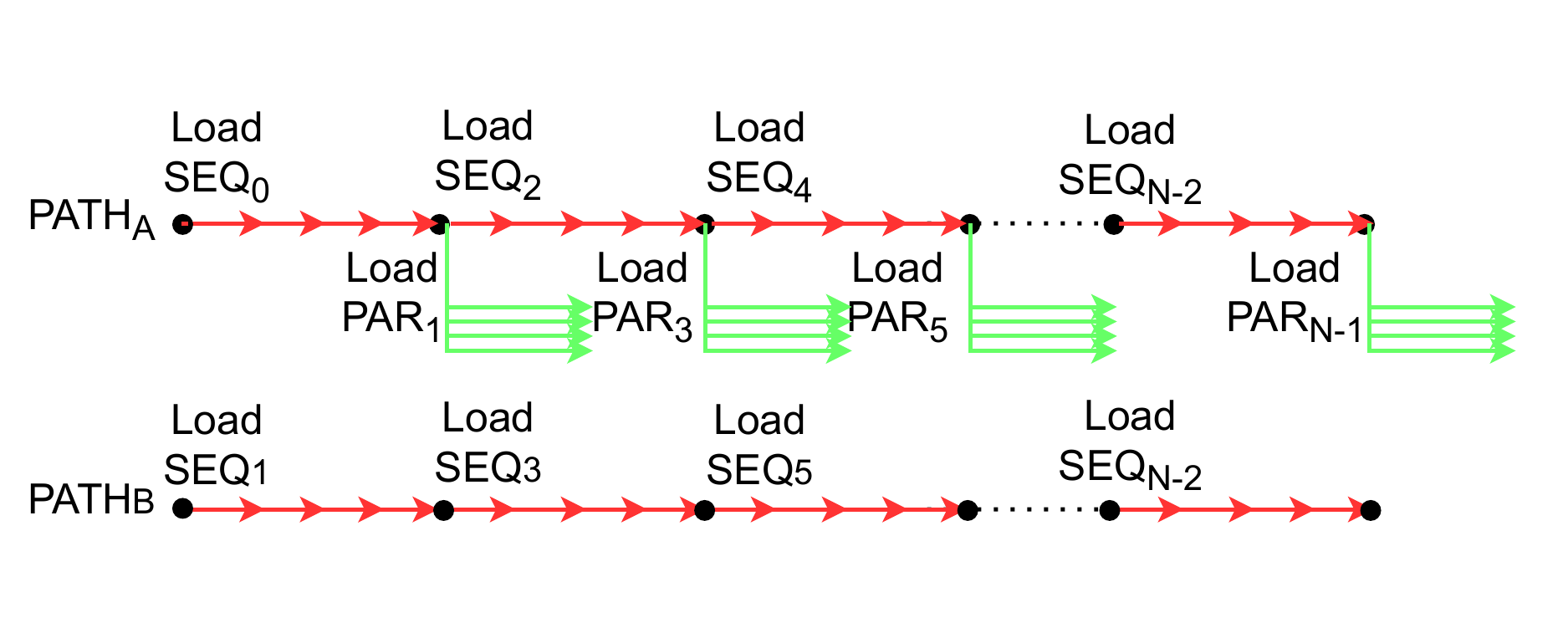}
\label{fig:path_nointerfere}
}
\captionsetup[subfloat]{farskip=-4pt,captionskip=-10pt}
\subfloat[The execution timeline of both paths when $Path_A$ starts earlier. The blue arrows indicates that the previous load evicted one data in the eviction set from that cache set. Thus, the loading time of $SEQ_i$ in $Path_B$ is longer than that in $Path_A$]{
        \includegraphics[width=\columnwidth]{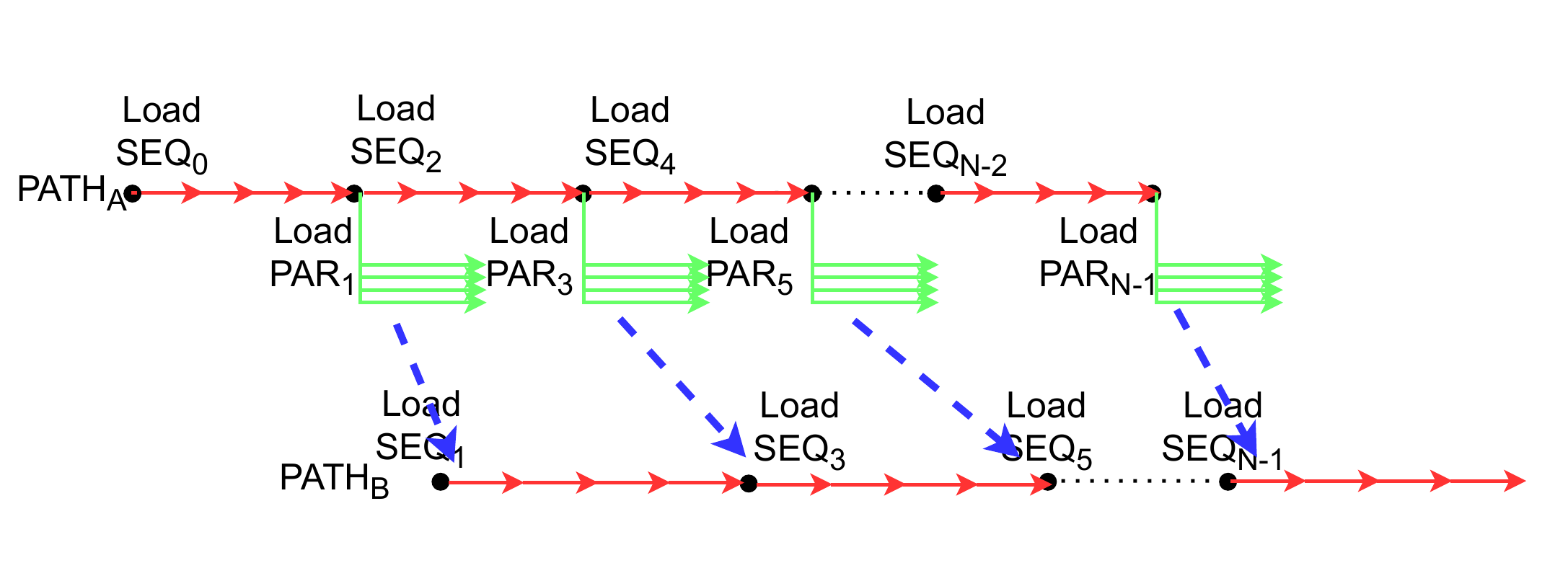}
    \label{fig:path_norepeat}
}
\captionsetup[subfloat]{farskip=-4pt,captionskip=1pt}
    
\subfloat[$Path_B$ prefetches for itself, to allow the magnification process to reuse the L1 cache sets cyclically and indefinitely.]{
        \includegraphics[width=\columnwidth]{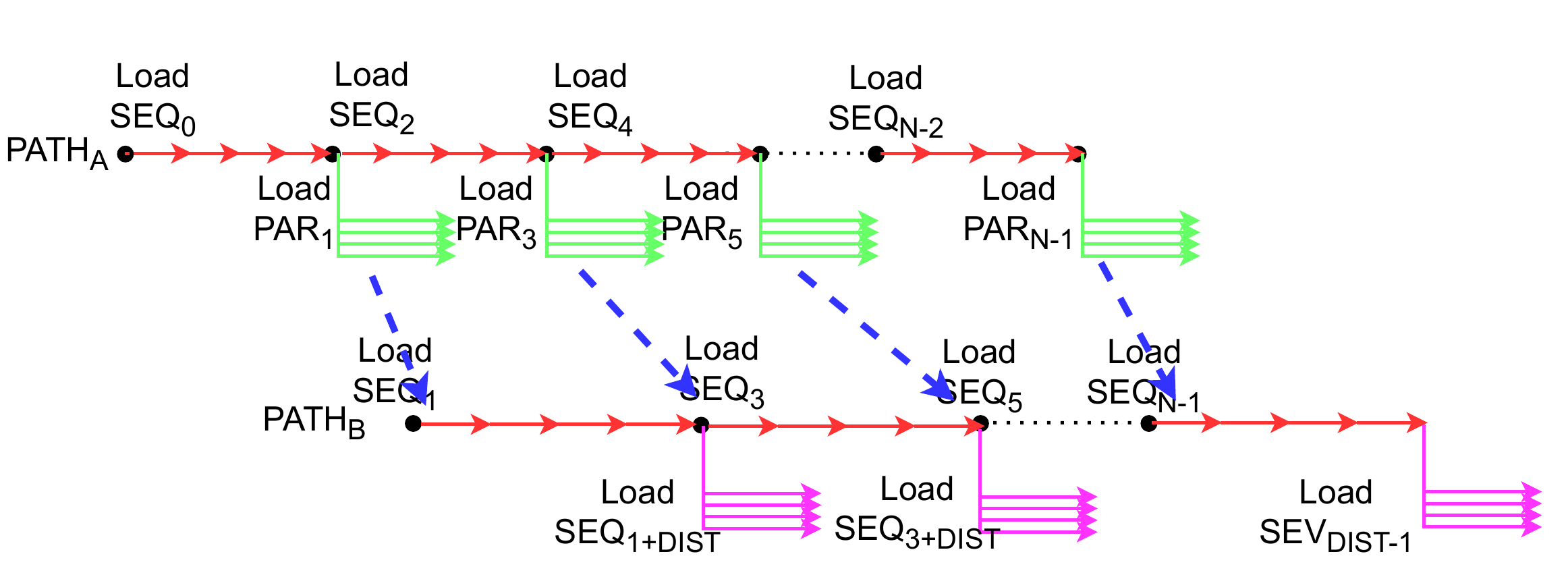}
    \label{fig:prefetch}
}
    
    \caption{The arbitrary-replacement-policy magnifier gadget. The magnifier is itself made up of two racing gadgets, $Path_A$ and $Path_B$, still both within the same thread. Different cache interference, based on whether one starts before the other, cascades into a succession of interference across the entire eviction set. Further, by allowing $Path_B$ to prefetch for itself in parallel with its loading of the sequential $SEQ_i$, the chain reaction can be continued indefinitely.}

\end{figure}

\subsection{Arbitrary Replacement Gadget for P/A Input}\label{sec: Arbitrary Replacement Gadget}

To demonstrate that changing the replacement policy will not eliminate this exploit, we introduce another \mgsing{} for caches with an arbitrary replacement policy within each set. 
 This gadget also utilizes ILP, using a chain reaction to amplify the timing difference of probabilistic state changes across many sets, rather than only converting the timing difference into a deterministic cache state change.

The basic strategy involves the magnifer gadget \textit{itself} becoming a racing gadget, with two paths that exhibit no contention when closely aligned in execution time, and heavy contention when misaligned (e.g. due to the unavailability of an input delaying one path). Once these paths become misaligned, each repetition stays misaligned, multiplying the delay for each stage to complete.

To simplify the demonstration, in the following we focus on an L1 cache with 64 sets, 8 ways and a random replacement policy~\cite{armrandomreplacement}.

The attacker first prepares the initial cache state across N (N $<$ 64) chosen sets, with W ways per set. The number N is linear to the final timing difference achieved, limiting the magnification achievable through sets alone, but we also develop a prefetching mechanism to further magnify this timing difference indefinitely. Two paths, $Path_{A}$ and $Path_{B}$, are generated in the attacker's code. We define both $SEQ_i$ and $PAR_i$ to each be subsets of addresses within the
eviction sets of the \emph{i}th cache set, without overlap between them. $PAR_i$ should be sized such that bringing its elements into the cache should evict at least one member of $SEQ_i$.

Before $Path_{A}$ or $Path_{B}$ start, we fill the N chosen sets with data from $SEQ_i$, ensuring that every element of $SEQ_i$ is within the cache\footnote{Even with random replacement, this initial state can be achieved through repeatedly accessing $SEQ_i$, provided $SEQ_i$ is smaller than the set size.}. The memory-access pattern in $Path_{A}$ and $Path_{B}$ is shown in figure \ref{fig:path_nointerfere}. When i is odd, $SEQ_i$ is accessed by $Path_{B}$. When i is even,  $SEQ_i$ is accessed by $Path_{A}$, followed by $PAR_{i+1}$. Arrows in figure \ref{fig:path_nointerfere} represents the data dependence that partially restricts the actual access order during runtime. For example, memory accesses within $SEQ_{0}$ are executed sequentially, while the accesses within the $PAR_{1}$ that immediately follow can be issued after the last access in $SEQ_{0}$, in parallel with each other. This access pattern ensures that the critical path of both $Path_{A}$ and $Path_{B}$ is only constructed from accesses in $SEQ_i$, as the $SEQ_i$ accesses will all hit in the cache, and so both paths should finish at approximately the same time when there is no interference between them.

By contrast, consider the timeline shown in figure \ref{fig:path_norepeat}, when $Path_{A}$ starts earlier than $Path_{B}$. By the time $Path_{B}$ starts loading $SEQ_{1}$, $Path_{A}$ has already accessed memory from $PAR_{1}$, which (potentially randomly) evicts elements in $SEQ_{1}$. Thus, one or several cache misses will occur when accessing $SEQ_{1}$. While the precise number of misses depends on the replacement policy, this will then delay the start of $SEQ_{3}$, which means that $Path_{A}$'s $PAR_4$ will create misses in $Path_{B}$'s $SEQ_4$. This delay accumulates over multiple rounds, with each round with a successful miss increasing the delay.

\subsubsection{Path prefetching}

As the number of cache misses induced in figure \ref{fig:path_norepeat} is linear to the number of cache sets, the amplification rate is limited, and so timers of coarser granularity would mitigate the attack mechanism. To succeed in spite of these, we can reuse the finite number of cache sets to generate theoretically unlimited timing difference, by generating a \textit{cycle of prefetches} \shepherd{(looping through $SEQ_i$)} within the paths to reformulate the initial conditions that caused the timing difference.

Specifically, because the initial cache state is destroyed after being accessed by $PAR_i$ from $Path_{A}$ in both the case of figure \ref{fig:path_nointerfere} and \ref{fig:path_norepeat}, we add parallel prefetch instructions on $Path_{B}$'s non-critical path with a prefetch distance DIST, as is shown in figure \ref{fig:prefetch}. These may be either software prefetch instruction or standard loads -- neither will block the out-of-order pipeline, and so neither will affect the execution time of $Path_B$'s critical path, save for the intended interference\footnote{\shepherd{Note that a software prefetch, especially if marked as non-temporal, might be inserted into ways that are easier to be evicted, making the timing difference easier to accumulate~\cite{leakyway}.}}.  This prefetching prepares the initial state for sets that will be later accessed by $Path_{B}$.

\subsubsection{What if $Path_{A}$ runs too far ahead of $Path_{B}$?} As there is no data dependence between the two paths, in figure \ref{fig:path_norepeat} and \ref{fig:prefetch} $Path_{A}$ could execute far ahead of $Path_{B}$ \shepherd{as time goes by} since there is no miss in its critical path. Although this could cause a processor stall as instructions from $Path_{A}$\shepherd{ and $Path_{B}$} fill the reorder buffer, it has little effect on the final timing. This is because the slower path, $Path_{B}$, continues execution during the stall, and $Path_{A}$ will still run ahead of $Path_{B}$ after the stall ends. \shepherd{Likewise, $Path_{A}$ starting too early has the same effect as both starting at the same time; due to limited ROB capacity, $Path_{A}$ cannot run ahead enough in this case to cause interference from different stages.}

\subsubsection{How many number of accesses should be included in each $SEQ_i$ and $PAR_i$?}

Larger sizes of $SEQ_i$ and $PAR_i$ will bring higher chance of cache misses, provided $SEQ_i$ fits in the cache.
We found that for a random replacement policy, with $SEQ_i$ set to 6 accesses (three quarters the associativity), 5 addresses in $PAR_i$ was sufficient to provide at least 1 cache miss in $SEQ_i$ with a 96\% chance, with larger values of either increasing the chance to near certainty. A failure to achieve a cache miss, beyond the first round, does not cause the attack to fail: it only causes a single round to not add any further delay. The subsequent round will still be delayed if any previous rounds were, and so will still be able to magnify the delay further.

\subsection{Arithmetic-Operation-Only Gadget for P/A Input}\label{sec: arithmetic}

\begin{figure}[t]
\captionsetup[subfloat]{farskip=-4pt,captionskip=-10pt}
\subfloat[The timeline of two paths when no interference occurs.]{
        \includegraphics[width=\columnwidth]{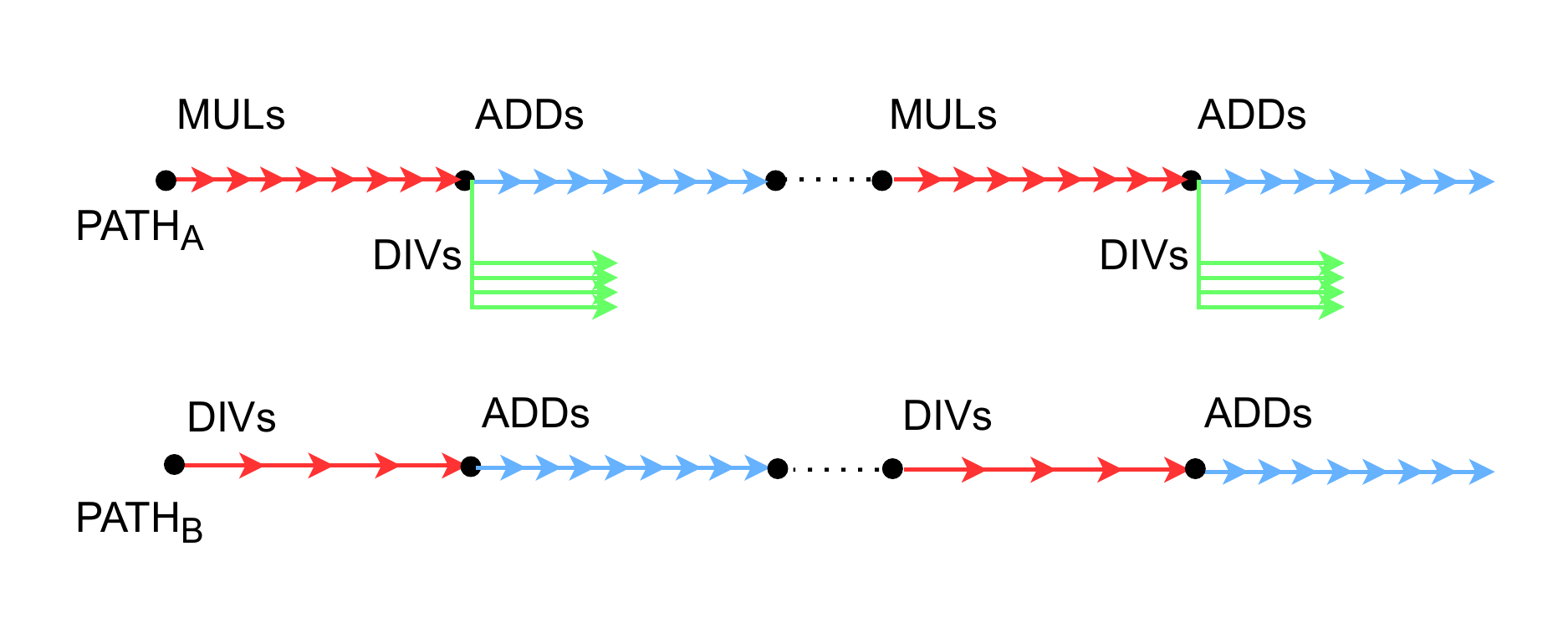}
\label{fig:arith_nocontention}
}
\captionsetup[subfloat]{farskip=-4pt,captionskip=-5pt}

\subfloat[The execution timeline of both paths when $Path_A$ starts earlier. The sequential DIVs in $Path_B$ start to contend with the parallel DIVs in $Path_A$, causing time displacement.]{
        \includegraphics[width=\columnwidth]{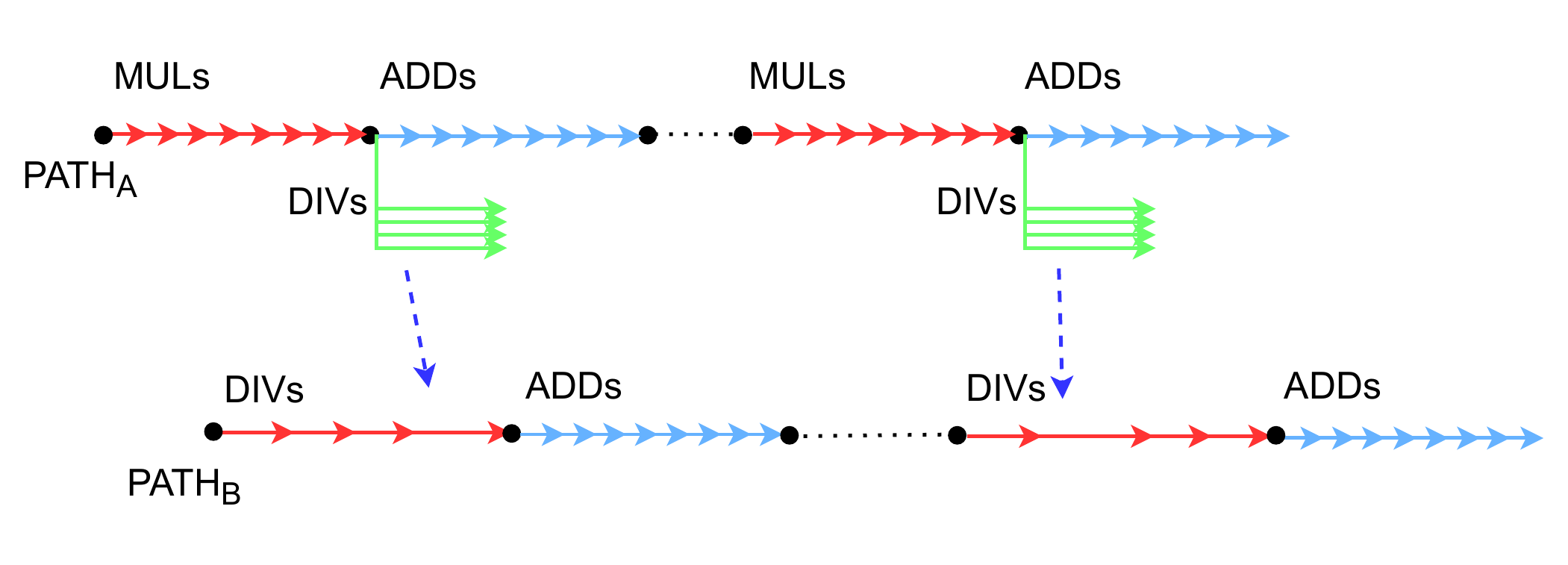}
    \label{fig:arith_contention}
}
    
 \caption{The arithmetic-operation magnifier gadget.}
 \label{fig:arithmetic}
\end{figure}

Although it is possible for previous \mg{} to succeed on cache-based channels, we provide a further gadget that makes no use of the cache whatsoever.
Since this gadget is composed of arithmetic operations, it eludes any arbitrary form of cache defence.

Similar to the gadget in section \ref{sec: Arbitrary Replacement Gadget}, chain reaction and ILP are also exploited in this gadget by composing two paths, $Path_A$ and $Path_B$. The difference is that sequential accesses are replaced by sequential arithmetic operations, and the contention in cache capacity is replaced by contention in not-fully-pipelined functional units (e.g. dividers). \shepherd{This magnifier is somewhat inspired by SpectreRewind~\cite{spectrerewind} and Speculative Interference~\cite{interference}, which use arithmetic-unit contention to generate Spectre side channels. However, here we repurpose such contention to generate much larger timing differences from a state generated via racing gadget: thus rather than a single timing difference, we must spawn a chain reaction.}

As is shown in figure \ref{fig:arithmetic}, $Path_A$ is constructed by chained integer multiply operations (MULs), chained integer add operations (ADDs) and parallel floating point divide operations (DIVs), while $Path_B$ is constructed from chained DIVs and ADDs. Without contention, the latency of each arithmetic operation is fixed. For simplicity, we assume $Latency_{DIV}$ = 9 cycles, $Latency_{MUL}$ = 3 cycles and $Latency_{ADD}$ = 1 cycles. 
The MUL and DIV operations serve as the racing stage, while the ADDs serve as a buffering stage.

When both paths starts execution at the same time in figure \ref{fig:arith_nocontention}, we would like the racing stage to finish at the same time. Since the latency of each DIV is three times that of a MUL, we set the number of MUL operations three times that of DIV. Therefore, the parallel DIVs after the racing stage in $Path_A$ will not delay the execution time of that in $Path_B$. In addition, we set the number of ADDs in each ADD chain to be the same as each other, and large enough so that the next racing stage will start at the same time, after all parallel DIVs have finished executing. As most CPUs can execute at least two ADDs at the same time, there is no interference within this buffer stage.

In contrast, figure \ref{fig:arith_contention} shows how contention cascades when $Path_B$ starts later. Since the execution time of their racing stages is the same, the parallel DIVs are issued before $Path_B$'s DIVs complete, and thus the two paths compete for resources. This results in $Path_B$ starting the ADD buffering stage later. As the number of ADDs is the same in both paths, $Path_B$ will then propagate this increased delay into the next racing stage against $Path_A$, which will grow with successive stages.

\subsubsection{What will happen if $Path_A$ runs too far ahead of $Path_B$?}
It is possible that $Path_A$'s parallel DIVs finish execution even before $Path_B$'s racing stage starts. In this case, $Path_B$ will no longer suffer from the DIV unit's contention, and the timing difference fails to increase. To solve this problem, we increase the number of operations in the racing stage, so that the limited capacity of the Reorder buffer (ROB) will stop this from occurring. Specifically, when $Path_B$ has not started, no MUL operation in the same racing stage can release its ROB entry. As long as the number of MUL operations exceeds the ROB capacity, the processor will stall before any parallel DIV can be issued, and can only continue executing in $Path_A$ after $Path_B$ also starts its racing stage. This stall only delays $Path_A$, so the timing difference will not be affected, as $Path_B$ is the critical path whose timing is observed.
\section{Attacks}

Here we demonstrate the utility of \name{}, by showing that simple repetition attacks alone do not typically give a timing difference that can be measured, but racing gadgets can fix them. Higher quality, faster bit-rate timers can be created by combining racing gadgets and magnifier gadgets, to allow extremely precise timing (to the nanosecond granularity) to be achieved even with the attack only being repeated once. Finally we use \name{} to construct a novel backwards-in-time Spectre attack~\cite{spectre}, demonstrate its efficacy even in JavaScript, and show how \name{} can be used as a drop-in replacement for SharedArrayBuffer timers~\cite{fantastictimers} by using them to construct LLC eviction sets.

\textbf{Processor details} We evaluate on an Intel i7-8750H Coffee Lake processor, though we reproduced similar results when we tried them on an AMD Ryzen 5900HX. The Intel system has 6 physical cores running at 2GHz. Each core has a private 32KB L1 Instruction Cache, 32KB L1 Data Cache, a 256KB L2 cache and a 64-entry L1 TLB. All cores share a 9MB L3 cache and a 1536-entry L2 TLB. The DIVSD instruction has a latency of 13-14 cycles based on the operand content, and 4-cycles reciprocal throughput~\cite{inslatency}. 

\shepherd{Nothing we implement is particularly microarchitecture-specific: our experimental artefact works across all Intel systems we tested out-of-the-box, and only the Prime-And-Scope attack (section \ref{ssec:llc}) fails on AMD systems, due to not having support for non-inclusive last-level caches within the underlying side-channel attack (rather than the Hacky Racer). Fundamentally the core needs to be out-of-order (i.e. any contemporary processor except the LITTLE part of Arm's big.LITTLE cores), and some of our individual attacks (but not all Hacky Racers) require either a set-associative cache or some non-fully-pipelined execution units. All of these are ubiquitous. In general we expect some attack-core combinations will require a simple profiling stage within the attacker's JavaScript.}

\subsection{Repetition Gadgets with Racing Gadgets}\label{sub: naive repetition}

\begin{figure}
    \centering
    \captionsetup[subfloat]{farskip=2pt,captionskip=-2pt}
    \subfloat[Time stack of the full execution of a basic repetition gadget. Without the use of racing gadgets, there is no discernible timing difference across the entire attack, as the cache miss that is saved in one path in the victim load stage is replicated instead in the attacker reload stage.]{
    \includegraphics[width=0.85\columnwidth]{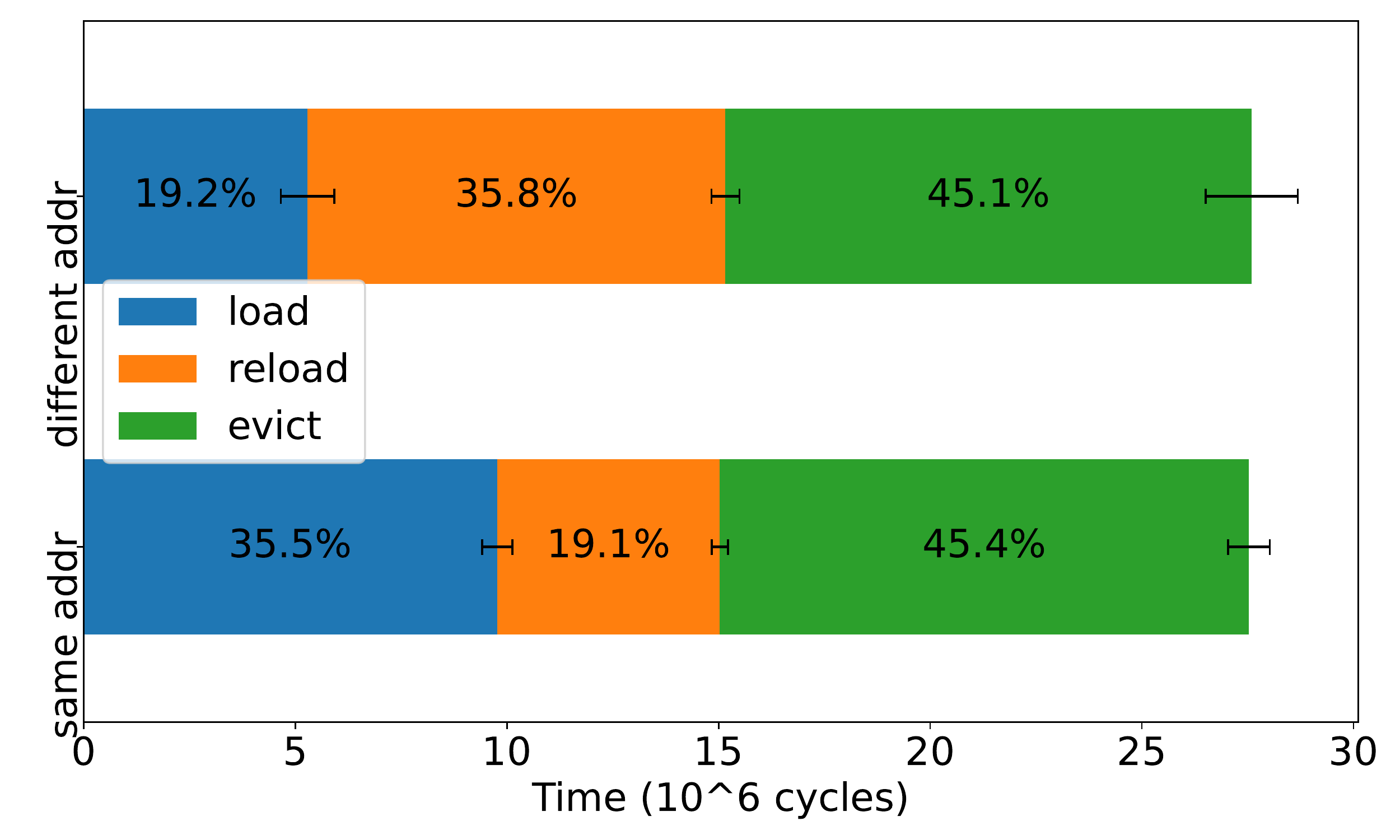}
    \label{fig: Time stack of repeated execution fail}
}

 \subfloat[Time stack of the full execution of an optimized repetition gadget, with a racing gadget used to make the load stage constant-time. The percentages shown are all normalized to the total runtime of the same-address case.]{
    \includegraphics[width=0.85\columnwidth]{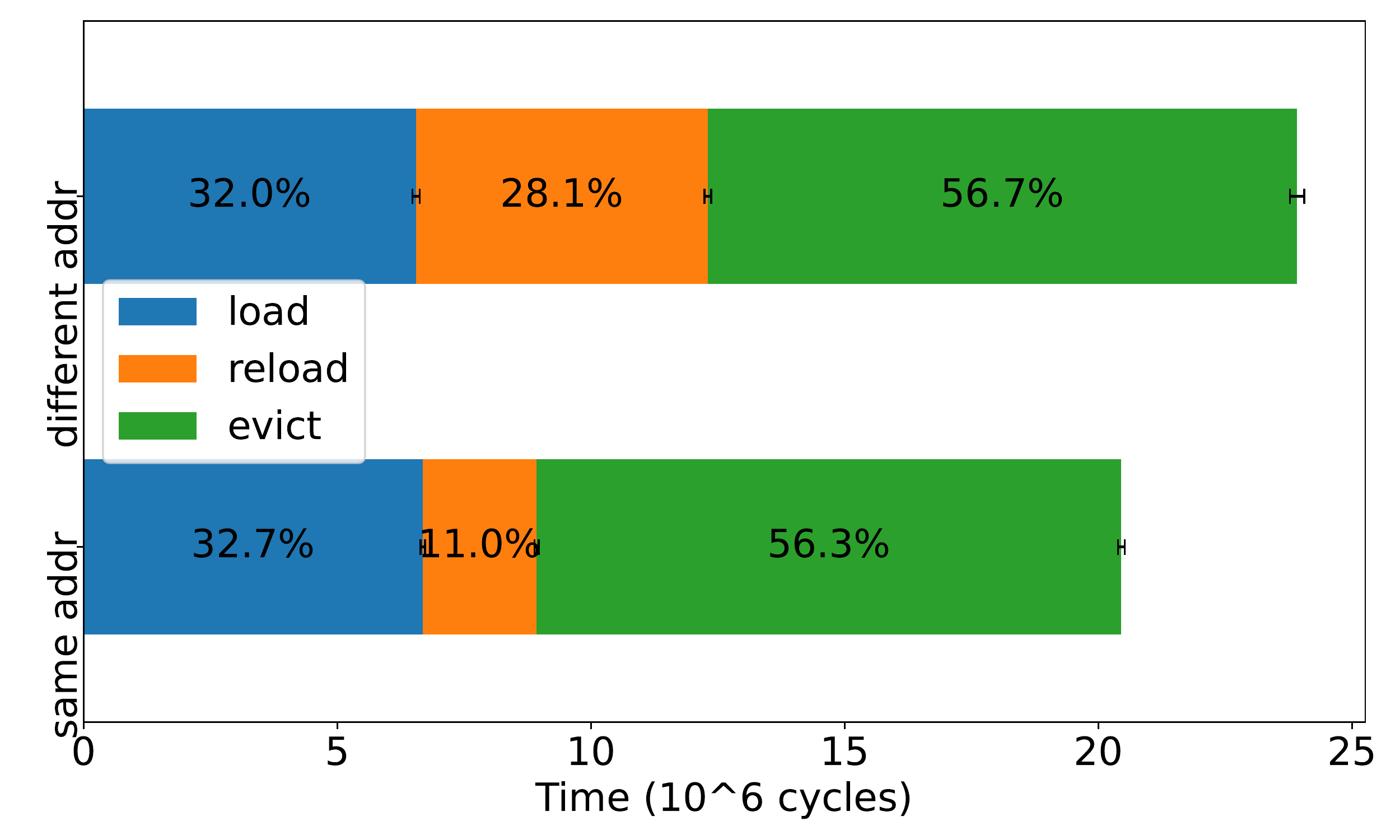}    \label{fig: Time stack of repeated execution succceed}
}

\caption{Repetition gadgets need to use racing gadgets in order to show a discernible timing difference.}

\end{figure}

Here we first give an example of how timing difference fails to accumulate by simple repetition~\cite{heretostay} of the \textit{flush+reload}~\cite{flushreload} process.  \shepherd{This is an arbitrary choice of attack, chosen to demonstrate that, counter-intuitively, a simple repetition can fail in some circumstances due to attack setup time, unless magnifiers are used to generate coarse timing difference instead.} 

The load stage will either access the same, or a different, address as the reload stage, which will be evicted in the flush stage. As is shown in figure \ref{fig: Time stack of repeated execution fail},  although the timing difference accumulates on the reload stage as expected, the loading stage has an opposite timing difference, eliminating the overall timing difference. 

This makes such a simple repetition ineffective as an arbitrary timing gadget; however, \name{} present a solution. We put the load stage into one path of a racing gadget, while the other path's execution time stays constant and always costs more time than the load stage. In this case, the timing difference from the load stage is hidden and the timing side-channel reappears in the total run-time, as is shown in figure \ref{fig: Time stack of repeated execution succceed}.

McIlroy et al.~\cite{heretostay} were able to use a repetition gadget to generate Spectre attacks\footnote{The load stage of a Spectre v1 gadget happens inside a misspeculated branch, whose return to correct execution is dictated by the branch-calculation time, which must be longer than the load access. This makes such a construction a racing gadget natively.}, but the leakage rate is 10B/s, 50$\times$ lower than the magnifier-based Spectre attack in section \ref{sec: JS backward-in-time}. This is because most execution time will be timing invariant, and to accumulate a large enough difference from the change in the reload stage alone takes significant time, impacting the bit rate. By combining with a magnifier gadget, we can make almost the entire execution time variant on the secret, increasing bit rates as a result.

\shepherd{This means that the combination of racing and magnifier gadgets are more capable than repetition gadgets alone -- as well as providing higher bit rates by making more of the execution time timing-variant: repetition alone is not always sufficient to generate a coarse time signal from an initial fine-grained timing signal. Still, racing gadgets alone are sufficient to improve a repetition gadget's signal-to-noise ratio, if not achieve high bit rates, by hiding the timing-variant setup time within a timing-invariant racing gadget.}

\subsection{Racing-Gadget Granularity}
\begin{figure}
    \centering
    \includegraphics[scale=0.5]{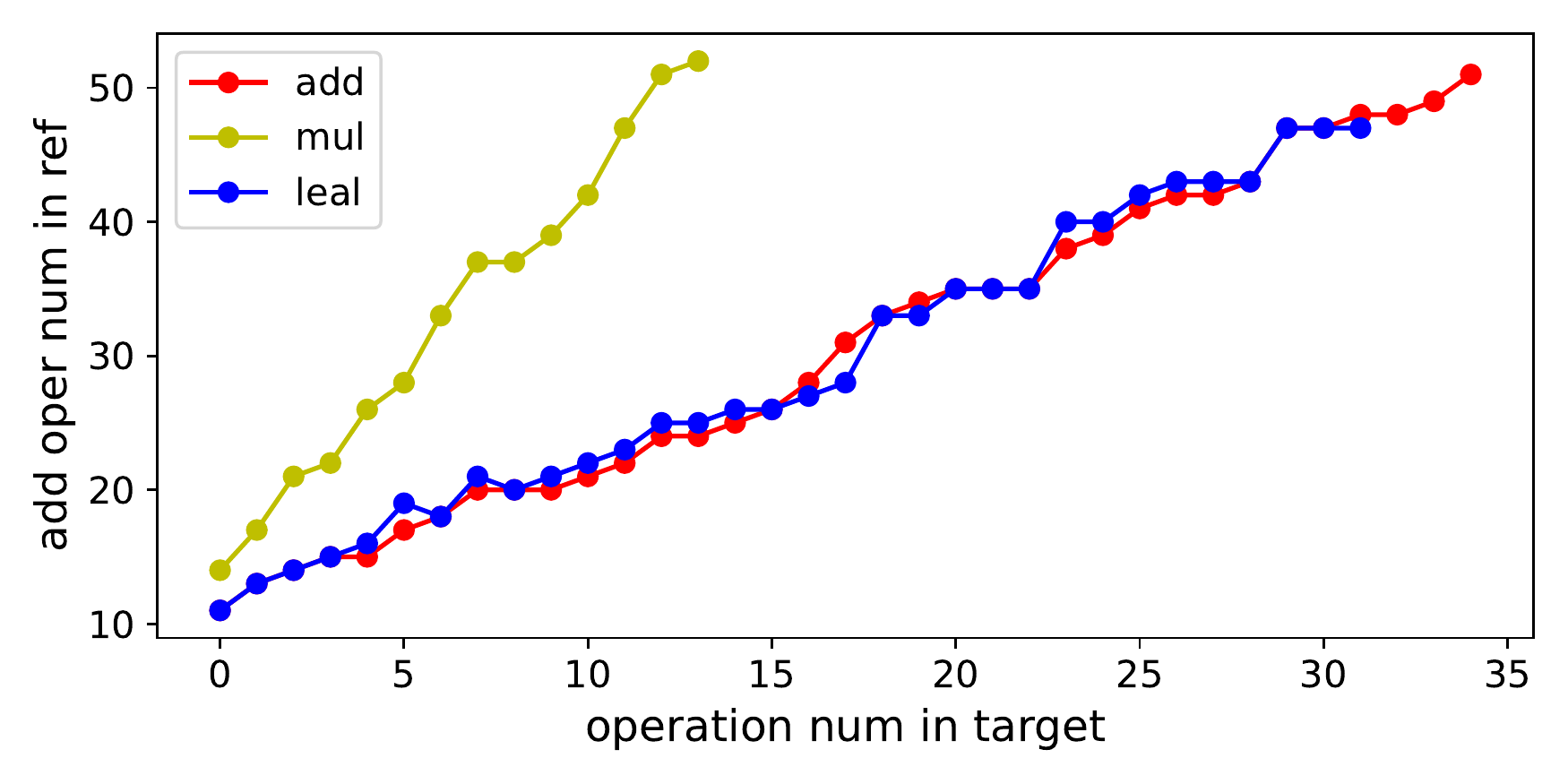}
    \caption{Target operation measured by reference path composed by ADDs. The granularity is the maximum consecutive points that are indistinguishable on the Y axis.}
    \label{fig: add as reference path}
\end{figure}

\begin{figure}
    \centering
    \includegraphics[scale=0.5]{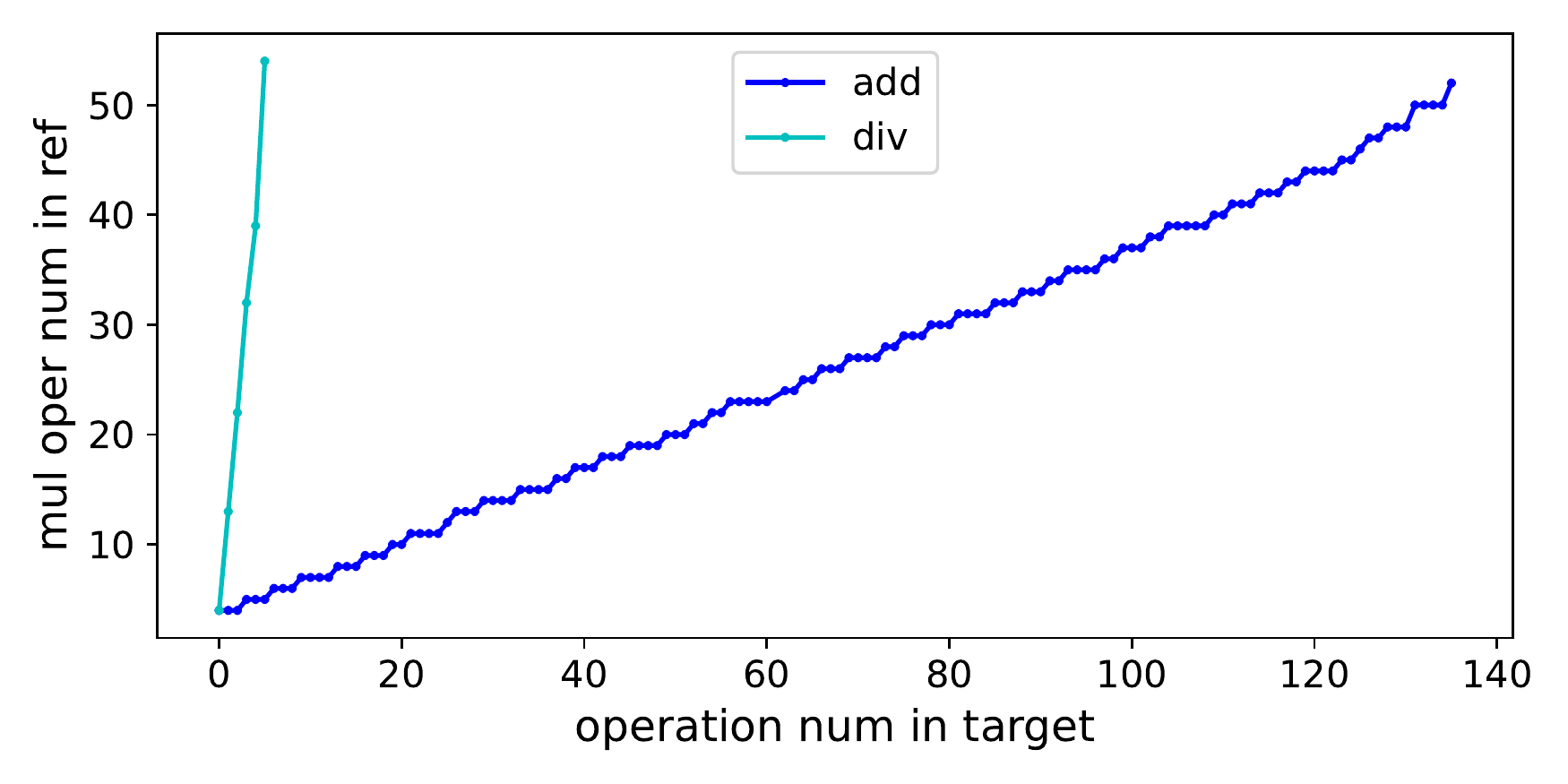}
    \caption{Target operation measured by reference path composed by MULs.}
    \label{fig: mul as reference path}
\end{figure}

To test the granularity of \rg{}, we choose the integer add operation to construct the reference path in a Transient P/A Racing Gadget, as each add operation only costs 1 cycle. We pick operations with various latency to construct the target path, and check whether our reference path can distinguish paths with different number of chained operations. The slope of each line is close to ratio between the latency of target operation and reference operation. The granularity is the maximum consecutive points whose Y value stays unchanged, indicating those two target paths cannot be distinguished by their corresponding operation number in the reference path.

For target paths composed of 1-cycle-latency chained operations as well, such as ADD and LEAL instruction shown in figure \ref{fig: add as reference path}, the granularity is 1 -- 3 operations. For MUL instructions (3-cycle latency), the granularity is 1 -- 2 operations. Therefore, the overall minimal granularity of \rg{} is 1 -- 6 cycles (0.5 -- 3ns). In this particular (Transient P/A) Racing Gadget, all instructions in the ref path are older than those in the target path (as the ref path is a misspeculated branch condition), thus the ROB capacity limits the length of the ref path to 54, which in turn limits the largest execution time that we can time to 54 cycles.

Using MUL operation, whose latency is longer than ADDs, we increase our timer's maximum timing threshold to around 140 ADD operations, as is shown in Figure \ref{fig: mul as reference path}. However, it sacrifices granularity to around 2 - 4 ADD operations. Still, it can distinguishes the number of DIV operations perfectly as DIV's latency is around 4 times of MUL's.

\subsection{Attack: SpectreBack in JavaScript}
\label{sec: JS backward-in-time}

\begin{figure}[t]
        \includegraphics[width=\columnwidth]{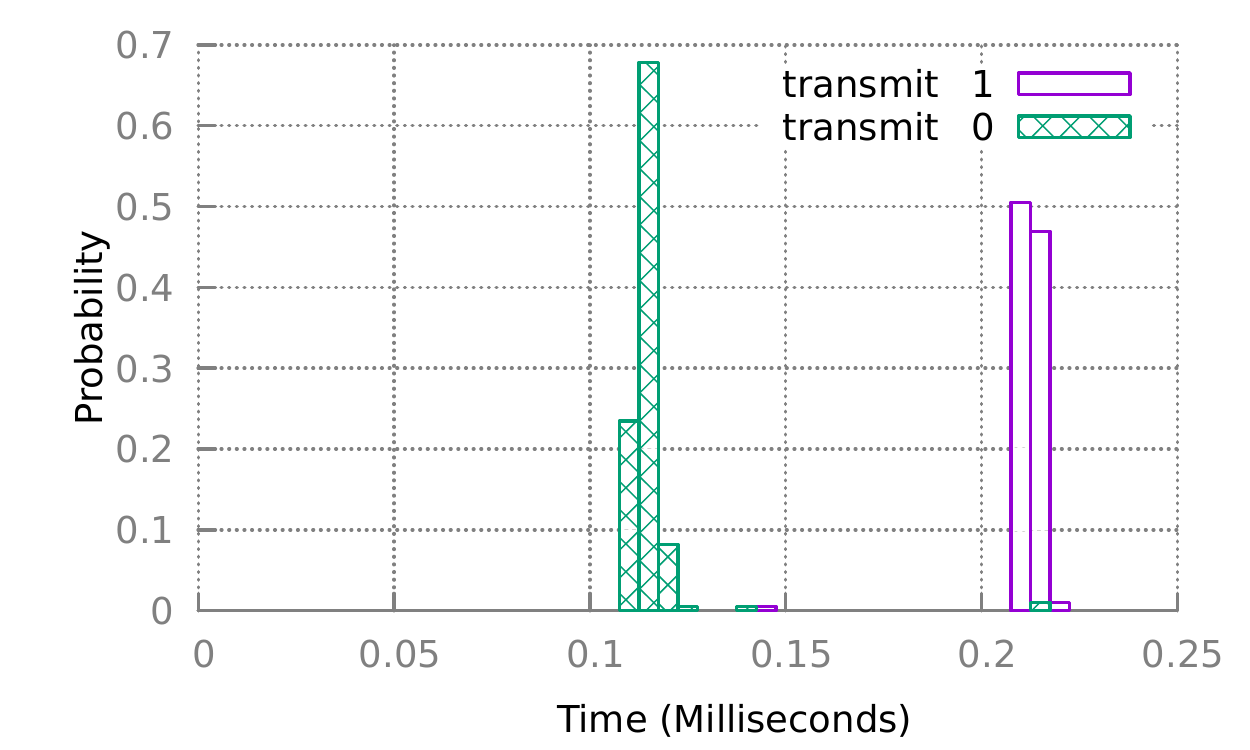}
    \caption{The execution-time distribution after the reorder magnifier access pattern is repeated 4000 times; there is still almost no overlap between the two transmissions.}
        \label{fig: reorder magnifier}

\end{figure}
Rollback-based~\cite{cleanupspec} and other Spectre mitigations that only clean up the effects of misspeculation once it has happened~\cite{muontrap,ghostloads,gonzalez2019replicating} (systems that do not implement strictness order~\cite{ghostminion}) were broken by attacks~\cite{spectrerewind, interference} that transmit the timing effect via contention. Here we propose a different method of breaking such systems, by transmitting timing information \textit{backwards in time} to a Non-Transient Reorder Racing Gadget (section \ref{sec: non-transient rg}) via cache state. \shepherd{It is presented here not only as a user of the reorder racing/magnifier gadgets, but also because the attack \textit{itself} is generated via a custom racing gadget, to transmit it to a reorder racing gadget that exists before any transient execution is squashed.} We use the PLRU Magnifier Gadget for Reorder Input to extract the secret.

The example code for this gadget is shown in code listing \ref{box:c}. The attacker controls which instruction sequence (line 4 and line 6) completes first based on the value of secret data speculatively accessed at line 8. This timing difference can be converted into the relative access order of array[A] and array[B]\shepherd{, similar to a non-transient reorder racing gadget} . This order will then be served as the input for the PLRU Gadget for Reorder Input (section \ref{plru reorder input}) at line 10. Figure \ref{fig: reorder magnifier} shows the output timing difference of this Gadget when its access pattern is repeated 4000 times.

Similar to leaky.page~\cite{treelru}, we use this gadget to leak out-of-bounds array data in Chrome 88, and achieve a 4.3 kilobits/second leakage rate with over 88\% accuracy (more than ample to leak secret keys).

\begin{BOX}[ht]
\centering
\begin{lstlisting}
function SpectreBack() {
  SetInitialCacheState()
  // access array[A] in the last, array[OFFSET] second last
  temp1 = array[array[array[a]]]
  // access array[B] in the last, array[OFFSET + 0x100] second last
  temp2 = array[array[array[b]]]
  // Based on the value of array[x], either access array[OFFSET] or array[OFFSET + 0x100]
  if (x < array_size) {temp3 = array[OFFSET + (((array[x]&bitmask) >> bit_num) << 8)]}
  time_begin = performance.now()
  traverseMagnifierPattern()
  time_end = performance.now()
  return time_end-time_begin;
}
\end{lstlisting}
\caption{SpectreBack, a new cache-based Spectre backwards-in-time attack. Based on the secret that is speculative accessed at line 8, it will either access \textit{array[OFFSET]} to accelerate line 6, or access \textit{array[OFFSET+0x100]} to accelerate line 4. Therefore, a relative access-order state is left as an input to the magnifier gadget at line 10. \label{box:c}}
\end{BOX}

\begin{figure}
    \centering
    \includegraphics[scale=0.4]{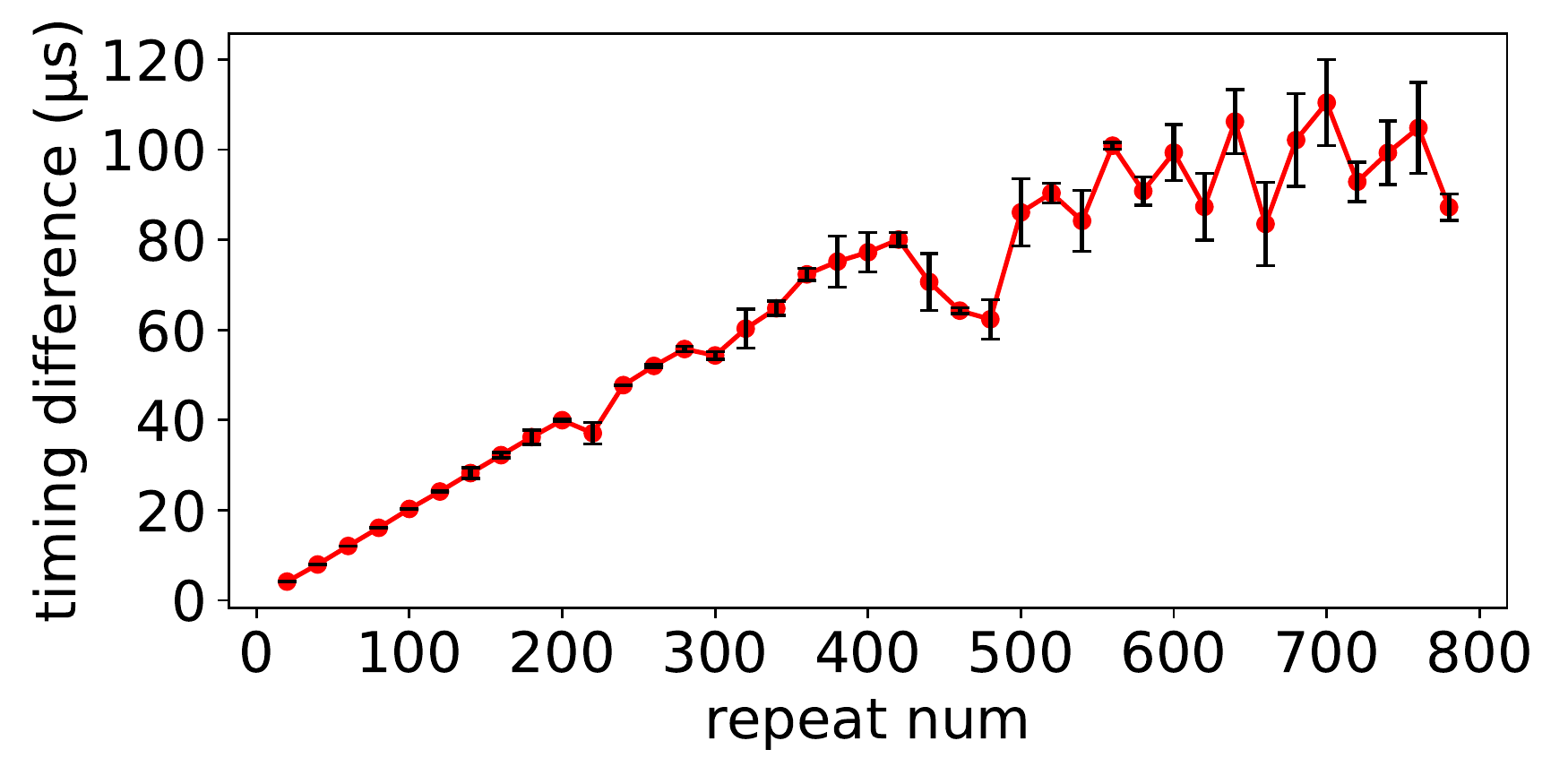}
    \caption{Timing difference magnified by the arbitrary-replacement gadget (\cref{sec: Arbitrary Replacement Gadget}) with cache-set reuse. Each set is used for contention once and reset to initial state by prefetching within one iteration.}
    \label{fig: sets reuse results}
\end{figure}

\begin{figure}
    \centering
    \includegraphics[scale=0.4]{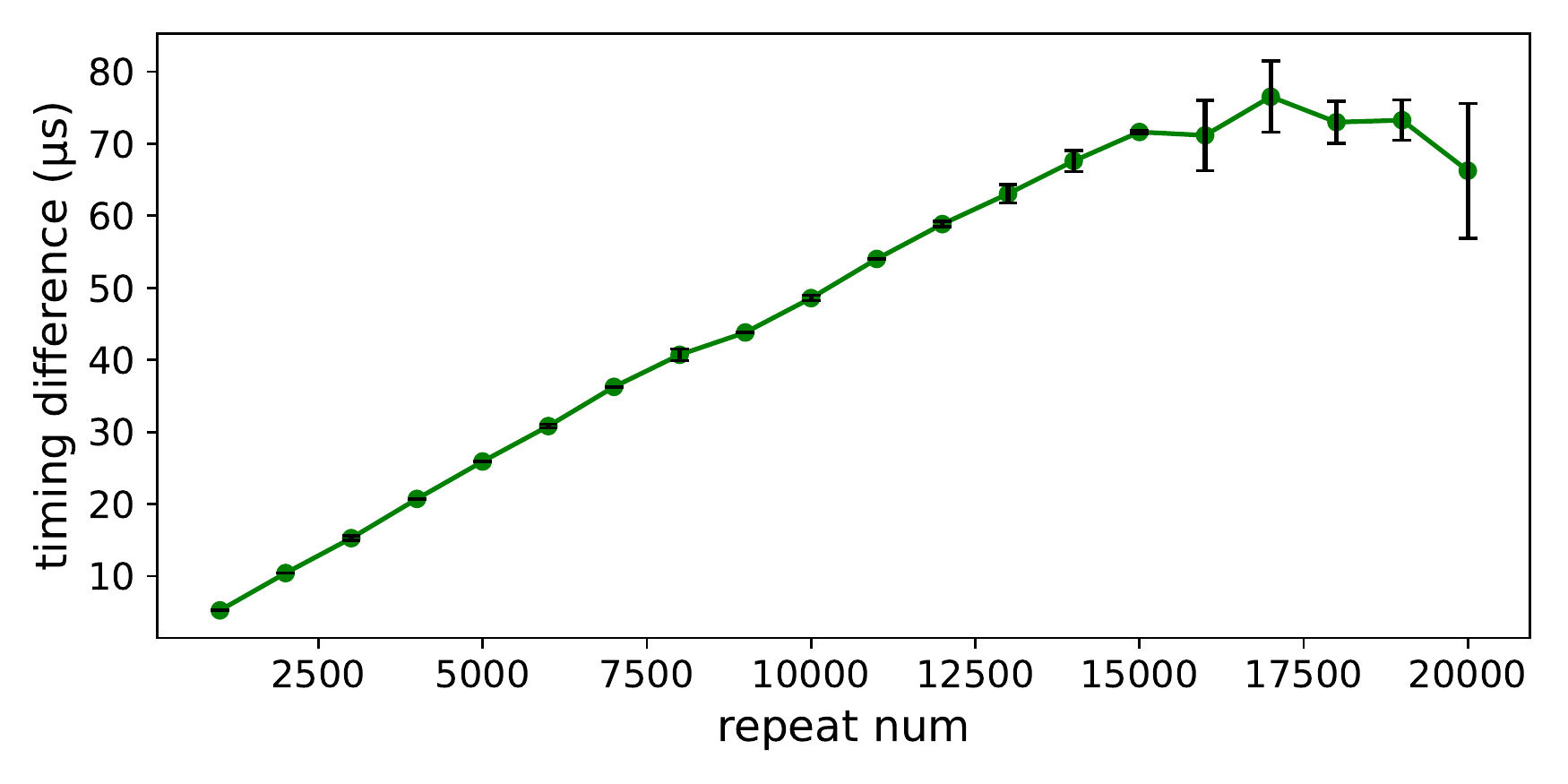}
    \caption{Timing difference magnified by arithmetic operations alone (\cref{sec: arithmetic}).}
    \label{fig: arithmetic magnifier result}
\end{figure}
\subsection{Attack: LLC Eviction-Set Generation}\label{ssec:llc}
To demonstrate that \name{} can be conveniently utilised by various attacks, we use a combination of the (P/A) Racing Gadget and PLRU Gadget for Presence/Absence (P/A) Input as the timer to generate eviction sets (EV), which allow the setup of various other attacks (\cref{sec: jstimer}). We use the EV profiling algorithm of Purnal et al.~\cite{primescope}, replacing just the timer. Since the timer in this algorithm only needs to distinguish between an \textit{L1 cache hit} and an \textit{L3 cache miss}, composing the reference path in the Racing Gadget with MUL operations can provide a fine enough granularity. We retained their 100\% success rate after the timer is replaced by ours. \shepherd{This is just one example that shows Hacky Racers can easily be used to replace existing SharedArrayBuffer timers in JavaScript -- and thus resurrects existing attacks even in hardened browsers.}

\subsection{Other Magnifiers}\label{sec:othermags}
Here we provide a PoC to show how our Arbitrary-Replacement and Arithmetic Gadgets (section \ref{sec: Arbitrary Replacement Gadget} and \ref{sec: arithmetic}) can also defeat coarsened timers, currently $5\mu s$, in browsers.

For the Arbitrary Replacement Gadget, we set half of the L1D cache's set number, 32, as the set number for each round. We set the prefetch distance as 22 iterations ahead. Figure \ref{fig: sets reuse results} shows that timing difference accumulates to $100\mu s$ as we repeatedly traverse through the sets; we also tried the arbitrary replacement gadget without prefetching, but as this was limited by the number of sets in the cache, it could only magnify up to 450 cycles (approximately 225ns).

For the Arithmetic-operation Only Gadget, as is shown in Figure \ref{fig: arithmetic magnifier result}, the timing different stops increasing at around a repeat number of 15000. This is because the total run-time approaches the interval of timer
interrupts (4ms), and since this magnifier is entirely stateless, it cannot keep accumulating timing difference once the pipeline is reset.

Both of these are more than ample to defeat current timer coarsening. Still, unlike the PLRU-based gadgets, whose magnification rates are almost arbitrary, these magnifiers are limited to rates close to the maximum ever granularity implemented in a browser (100ms~\cite{soklosttime}). In practice this is little impediment; both can be combined with repetition gadgets (increasing bit rate compared to repetition alone, by making more of the execution timing-variant, and avoiding any negative correlation with setup time masking the time difference), at the expense of needing the attack to be performed multiple times. Whether either or both can have their magnification rates improved, allowing the attack to be repeated only once even for arbitrarily coarse timers, is open future work.
\label{attacks}
\section{Potential Countermeasures}
\shepherd{Transient execution is not required for racing or magnifier gadgets: only the transient presence-absence (P/A) racing gadgets in section {\ref{sec: p/a racing gadgets}} (because they use transient execution directly) and the SpectreBack attack in section {\ref{sec: JS backward-in-time}} (because it is a Spectre attack, though it does break Spectre defences}~\cite{cleanupspec,muontrap,ghostloads,gonzalez2019replicating}\shepherd{ that do not guarantee Strictness Order~\cite{ghostminion}, and thus try to clean up misspeculation after it has occurred), use transient execution in any way, and thus only these can be guarded by Spectre defences}~\cite{DoM, cleanupspec, muontrap, ghostloads, ghostminion, gonzalez2019replicating, invisispecbug, stt, sdo, nda, safespec}\shepherd{, which are designed to eliminate the negative effects of transient execution. We use transient (P/A) gadgets in many of our examples because they are simple to understand: however, an attacker can easily change to use reorder gadgets instead.}

\shepherd{Take DoM~\cite{DoM} (delay-on-miss), a technique which delays cache misses in the L1 until they become non-speculative, and the non-transient reorder racing gadget as an example. The reorder gadget only relies on the relative cache insertion order of two load instructions. Both of these can be entirely non-speculative, and will still race due to instruction-level parallelism, and yet DoM marks them as being safe to execute in any order. To take another example, GhostMinion~\cite{ghostminion} might change the original L1 cache insertion order by by first inserting speculative load into the GhostMinion cache, the exact state that ends up being inserted into the L1 cache still ends up dictated by the out-of-order execution present in the reorder gadget\footnote{\shepherd{The strictness order presented in GhostMinion~\cite{ghostminion} does not guarantee the cache state of the L1 is always equivalent to an in-order execution. While it ensures that} \shepherd{a less speculative instruction cannot see out-of-order execution of a more speculative instruction, it does not guarantee the opposite, as that is not necessary to hide} \shepherd{transient execution. So a load at timestamp $21$ may validly be evicted from the GhostMinion by a load at timestamp $20$ if the latter happened afterwards in out-of-order execution -- meaning that only load $20$ ends up in the L1 cache after -- whereas if they were executed in program order, both may reach the L1 in sequence.}}.} \shepherd{The fundamental property is that Spectre defences} \shepherd{treat transient execution as the dangerous part; and indeed while} \shepherd{it is more dangerous than Hacky Racers alone (which leak timing information rather than secrets directly), they do not seek to hide or eliminate channels caused via instruction-level parallelism.}

\shepherd{In general, Hacky Racers can be defeated by either preventing the racing gadget's success in creating the state change, or the magnifier gadget's success in replicating it -- though the attacker may be able to rely on repetition rather than magnification in circumstances where this still provides a high enough signal-to-noise ratio (section \ref{sub: naive repetition}). Some of our magnifiers are limited in magnification capability, and could be defeated via further coarsening (section \ref{sec:othermags}), whereas others (the PLRU gadgets) are unlikely to be limited without removing any source of coarse-grained time completely, which is likely to be impossible in practice. We expect coarser-grained magnifiers not using PLRU to follow, and even our least effective magnifiers would still combine well with repetition in such an environment.}

\shepherd{One potential way to completely mitigate the cache-based reorder gadget we give in this paper is to guarantee cache state is always equivalent to an in-order execution. We leave the design, evaluation and overheads (and whether such overheads are feasible for deployment, given the existing overheads of Spectre defences without such guarantees) of such a scheme to future work. Likewise, our reorder magnifier is based on PLRU replacement, and so other policies will break this specific gadget, though we present others that translate over simply. But even if this cache-based gadget is mitigated, an attacker can then change strategy to transmit timing based on within-core contention -- where in the general case, assuring behavior equivalent to in-order execution is likely to require actual in-order execution.}

\shepherd{For run-time detection mechanisms~\cite{briongos2018cacheshield, alam2017performance, chiappetta2016real, payer2016hexpads, zhang2016cloudradar}, we expect racing gadgets to look so similar to normal out-of-order execution that they will be difficult to catch without very high false positive rates. Still, since magnifier gadgets rely on highly repetitive patterns, detection of at least high-bitrate channels may be feasible, but will require different schemes for every possible repeated pattern, and so attackers will keep changing strategy. Frequent L1 cache misses exists in PLRU gadget of section \ref{pa gadget} and Arbitrary Replacement Gadget in section \ref{sec: Arbitrary Replacement Gadget}, therefore the L1 cache miss counter could be utilized to as one input to such a detector, though only as a very weak classifier. Similar to the port contention attack by Rokicki et al.~\cite{portcontetionILP}, the Arithmetic-Operation-Only Gadget of section \ref{sec: arithmetic} executes long backend-bounded instruction (arithmetic operations) chain without misprediction, which makes the ratio of backend-bound execution divided by misprediction-bound execution also a potential parameter to detect this gadget. Besides, since this attack also requires precise instruction sequence construction like that by Rokicki et al.~\cite{portcontetionILP}, the reordering of instruction sequences from either browser's optimization or software-diversification}~\cite{thwartingviaswdiversity, raccoon}\shepherd{ can affect the efficiency of Hacky Racers, and software analysis~\cite{microwalkci} within JavaScript compilers may be able to pick up attacks that are less well obfuscated. Still, in some scenarios attackers may have enough control to manually construct the instruction sequence or implement multiple backup sequences to overcome any such analysis. Again, we leave this for future study.}

\shepherd{For browser security, where the main threats are from making Spectre and fingerprinting attacks simpler even without SharedArrayBuffer timers, cross-origin isolation policies~\cite{coop} (as opposed to site isolation alone which was attacked by spook.js~\cite{spook.js})}\shepherd{ move many targets outside the address space, although implementation requires manual effort to avoid breaking compatibility, and so adoption is not yet universal. A similar process-level isolation policy~\cite{dynamicprocessisolation}} \shepherd{ is needed in Cloudfare, otherwise Hacky Racers will make the Spectre-attack proposed by Martin et al.~\cite{dynamicprocessisolation} much easier. We also believe that Tor, which currently disables \textit{SharedArrayBuffer} even with cross-origin isolation~\cite{Torsharedarraybuffer}, now gains very limited protection from this constraint, in the presence of Hacky Racers.}
\section{Discussion and Conclusion}

\name{} show that high-resolution timing information can be extracted via instruction-level parallelism, even in highly restricted JavaScript environments, and even without multiple threads~\cite{fantastictimers} or any language features that could realistically be disabled. We do not believe mitigation to be particularly realistic; out-of-order execution is vital for single-threaded performance (so systems will not be going in-order), removal of PLRU cache replacement~\cite{pseudolru} will only cause the attacker to change strategy, and while some of our gadgets could be eliminated through further coarsening, others work to almost arbitrary degree, and all can be amplified further by repetition. We also expect further magnifiers to follow. This demonstrates that timer coarsening has limited efficacy in any browser security model, and the removal of SharedArrayBuffer from sites without cross-site isolation~\cite{chromesharedarraybuffer} in Chrome and Firefox is insufficient to protect them from timing side channels. Threats will have to be dealt with, in future, by isolation~\cite{siteisolation} only.  \name{} may also impact other security models with restricted timing, such as the removal of user-privilege fine-grained timers on M1 processors and other ARM systems~\cite{pacman, armageddon}; we leave consideration of these to future work.

Our attacks also have implications for architects. While Spectre~\cite{spectre} showed speculative execution to be a fundamental source of information leakage, \name{} goes even further: even correct execution results in information leakage. Is any microarchitectural performance optimisation truly secure, given the right threat model?

\section{Acknowledgement}
We would like to sincerely thank ASPLOS 2023 Reviewers for your helpful feedback. We would also like to thank Thomas Rokicki for providing a modified version of Chromium.

\bibliographystyle{plain}
{
\footnotesize
\bibliography{references}
}
\end{document}